\documentclass[epj]{svjour}
\usepackage{graphics}
\usepackage{graphicx}
\usepackage{subcaption} 
\usepackage{mathtools}
\usepackage{slashed}
\usepackage{url}
\usepackage[normalem]{ulem}
\usepackage{enumitem}
\usepackage{mathrsfs}
\usepackage{float}
\usepackage{hyperref}
\usepackage{natbib}
\usepackage{xcolor}
\usepackage{bbold}
\newcommand{\orcid}[1]{\href{https://orcid.org/#1}{\includegraphics[width=8pt]{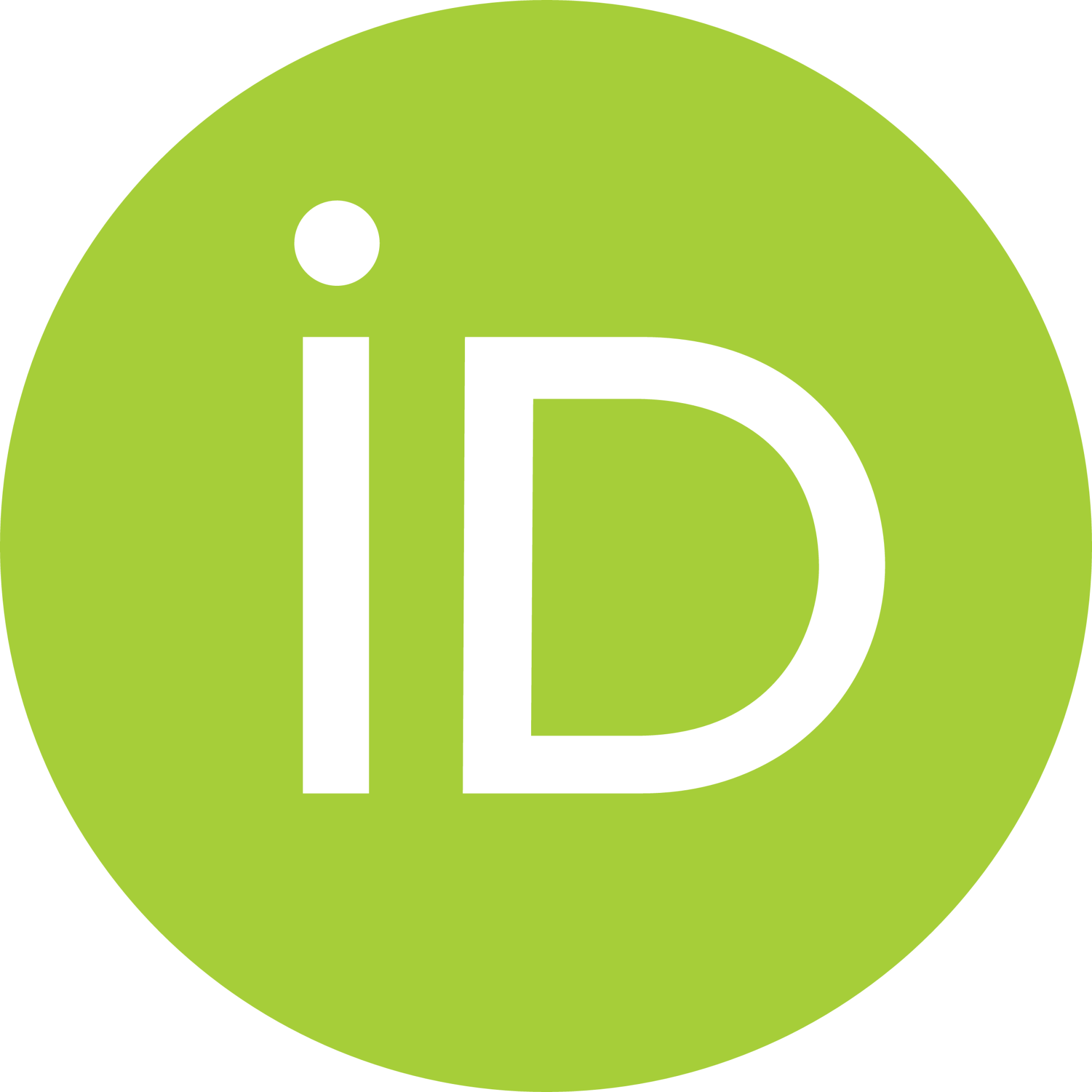}}}

\begin{document}
\title{Towards compressed baryonic matter densities: thermodynamics and transport coefficients}

\author{
	Anand Rai\orcid{0009-0005-4761-7918}
	\and
	Dani Rose J Marattukalam\orcid{0009-0006-8204-8148}
	\and
	Prasanta Murmu\orcid{0009-0006-8532-7679}
	\and
	Ashutosh Dwibedi\orcid{0009-0004-1568-2806}
	\and
	Rishabh Sharma\orcid{0009-0006-7751-0502}
	\and
	Sabyasachi Ghosh\orcid{0000-0003-1212-824X}
}

\institute{Department of Physics, Indian Institute of Technology Bhilai, Kutelabhata, Durg 491002, India}
\date{Received: date / Revised version: date}
%
\abstract{We study the thermodynamic and transport properties of hot and dense quantum chromodynamic matter expected to be produced in low-energy heavy-ion collisions, using three different effective quantum chromodynamic frameworks: the Nambu–Jona-Lasinio model, the chiral effective model, and the hadron resonance gas model. We briefly outline the theoretical formulation of thermodynamic quantities and transport coefficients within these approaches, where quarks are treated with effective masses in the Nambu–Jona-Lasinio and chiral effective models, and hadronic degrees of freedom are employed in the hadron resonance gas model. The transport coefficients are evaluated using the Boltzmann transport equation in the relaxation-time approximation. Following the theoretical overview, we present a comprehensive analysis of the behavior of these quantities as functions of the baryon chemical potential or net baryon density. The Lorenz ratio $\kappa/(\sigma T)$ is found to increase rapidly—indicating a strong violation of the Wiedemann–Franz law in the low-$\mu_{B}$ regime—while approaching the universal value at higher baryon chemical potentials or densities. The shear-viscosity-to-entropy-density ratio $\eta/s$ remains nearly constant at low $\mu_{B}$ but exhibits a gradual increase as $\mu_{B}$ grows. We also discuss the qualitative similarities of these trends with those observed in the electron–hole plasma of graphene, an emergent quasi-relativistic system characterized by massless energy–momentum dispersion. 
	\PACS{
		{}{}   \and
		{}{}
	} 
} 

\maketitle

\section{Introduction}

Understanding the thermodynamic and transport properties of hot and dense matter is essential for comprehending the evolution of strongly interacting systems in relativistic heavy-ion collisions. The equation of state, which relates the energy density to the pressure, along with the transport coefficients including shear viscosity, bulk viscosity, and electrical conductivity, serves as input to the hydrodynamic evolution of the fireball \cite{Heinz:2005bw,Gale:2013da}.

The calculation of these coefficients employs underlying microscopic theories that can be broadly classified into two approaches: the fundamental quantum statistical linear-response frameworks, such as Kubo-like approaches \cite{Ghosh:2014yea} and the kinetic description based on the Boltzmann transport equations \cite{DeGroot:1980dk}. In this work, we employ the latter approach of kinetic theory, and most of the following discussion will be based on this framework \cite{Gavin:1985ph,Chakraborty:2010fr}. The temperature dependence of thermodynamic quantities \cite{Chandra:2009jjo,Chandra:2011en,Bhattacharyya:2010wp,Bhattacharyya:2012rp,Bhattacharyya:2014uxa,Bhattacharyya:2015zka,Satapathy:2019jdw} and transport coefficients \cite{Ghosh:2014yea,Ghosh:2013cba,Abhishek:2017pkp,Ghosh:2014qba,Singha:2017jmq,Kadam:2015xsa,Kadam:2014cua,Deb:2016myz,Thakur:2017hfc,Shaikh:2024gsm,Mitra:2017sjo,Mitra:2018akk,Pradhan:2022gbm,Goswami:2025pnr,Sahu:2020mzo,Satapathy:2019jdw} \footnote{also see the articles cited in refs.\cite{Ghosh:2014yea,Chandra:2009jjo,Chandra:2011en,Bhattacharyya:2010wp,Bhattacharyya:2012rp,Bhattacharyya:2014uxa,Bhattacharyya:2015zka,Satapathy:2019jdw,Ghosh:2013cba,Abhishek:2017pkp,Ghosh:2014qba,Singha:2017jmq,Kadam:2015xsa,Kadam:2014cua,Deb:2016myz,Thakur:2017hfc,Shaikh:2024gsm,Mitra:2017sjo,Mitra:2018akk,Pradhan:2022gbm,Goswami:2025pnr,Sahu:2020mzo,Satapathy:2019jdw} } have been extensively studied, particularly at zero net baryon density  and has been successful in describing nearly baryon-free matter formed at the top Relativistic Heavy Ion Collider (RHIC) energies at Brookhaven National Laboratory, USA and Large Hadron Collider (LHC) energies at CERN.

However, with the upcoming Compressed Baryonic Matter (CBM) experiment at FAIR, Germany, and the Nuclotron-based Ion Collider fAcility (NICA) at JINR, Russia—where a finite baryon density is expected—microscopic calculations of thermodynamical quantities and transport coefficients of baryon-rich quark or baryonic matter will become important and timely research topics. A few studies in this direction can be found in Refs.~\cite{Albright:2015fpa,Albright:2015uua,Schaefer:2008hk,Ghosh:2018xll,Abhishek:2017pkp,Soloveva:2019xph,Soloveva:2020hpr,Fotakis:2021diq,Du:2019obx,Du:2022yok}; however, a comprehensive understanding of the baryon-density dependence is still lacking. The first principle lattice quantum chromodynamics (LQCD) calculations fall short in this regime of non-zero baryonic density due to the numerical sign problem, ascertaining the need for more extensive phenomenological studies in this regime. The present work aims to contribute to an extensive understanding of the thermodynamic and transport properties of CBM or NICA matter, where three different models - the Hadron Resonance Gas (HRG) model, the mean-field chiral effective model, and the Nambu–Jona-Lasinio (NJL) model- are considered.  

The success of the HRG model as an effective theory of hadronic matter relies on the fact that it reproduces the results of first-principle LQCD simulations with exceptional accuracy in the hadronic temperature regime \cite{Andronic:2012ut,Karsch:2003vd, Ghosh:2019fpx,Biswas:2022vat,Biswas:2024xxh}. At higher temperatures and densities, where chiral symmetry restoration is anticipated, the NJL model \cite{Klevansky:1992qe, Buballa:2003qv} captures crucial QCD features like dynamical mass generation and chiral symmetry breaking \cite{Klevansky:1992qe, Buballa:2003qv,Hatsuda:1994pi}. Using NJL model, Ref \cite{Ghosh:2015mda,Bandyopadhyay:2023lvk,Sasaki:2008um,Marty:2013ita,Islam:2019tlo} have estimated transport coefficients of quark matter. The chiral effective models, based on chiral symmetry breaking and the broken scale invariance of QCD, have been successful in explaining the in-medium properties of finite nuclei, nuclear matter, strange baryonic matter \cite{Papazoglou:1997uw, Papazoglou:1998vr,Mishra:2003tr, Zschiesche:2003qq, Mishra:2006wy}. The use of the chiral effective model as a quasiparticle quark model for the calculation of transport coefficients was introduced in \cite{Marattukalam:2024mef,Marattukalam:2025lpi}, where the temperature and density dependent constituent quark mass maps the medium effects of the system. Other similar works include the use of quark \cite{Kumari:2020mci,Chahal:2022syd,Singh:2024pwv,Singh:2025nri} and hadronic \cite{Kumar:2025rxj} chiral effective models to study the thermodynamic and transport properties at finite densities. Taken together, the HRG, NJL, and chiral models provide a complementary and comprehensive framework for investigating the temperature and density dependence of QCD matter across the entire phase diagram. Such studies are crucial for interpreting current results from RHIC and LHC, as well as for guiding future experiments at FAIR-CBM and NICA~\cite{Wilczek:2010ae}.

In this work, we investigate transport coefficients and thermodynamic observables in the light sector - specifically for a three-flavor system - using the HRG, NJL, and chiral effective models at finite densities. As the accepted benchmark and to provide a quantitative comparison of quantities obtained using different effective models, we first discuss their temperature dependence at zero net baryon density. The primary focus of this article is to provide a coherent investigation of the density dependence of the thermodynamic and transport properties of QCD matter, with emphasis on the lower-energy regime of experiments like CBM and NICA. 

The paper is organized as follows. In Section~\ref{sec:formalism}, we outline the theoretical framework. Subsections~\ref{NJL}-\ref{HRG} present the details of the three effective models used -- the NJL model, the chiral effective model, and the HRG model. Subsection~\ref{thermo-transport} reviews the thermodynamic quantities and transport coefficients within kinetic theory. The results are presented in Section~\ref{sec:results}, followed by a summary in Section~\ref{Summary}. 

\section{Formalism}
\label{sec:formalism}

\subsection{Nambu--Jona-Lasino (NJL) Model}\label{NJL}

In this section, we outline the NJL model, which effectively captures the mechanism of spontaneous chiral symmetry breaking in QCD, and derive the quasiparticle quark masses as functions of temperature and density of the medium. In the NJL model, the dynamical chiral symmetry breaking is realized by introducing a four-fermion and six-fermion contact interactions~\cite{Klevansky:1992qe,Buballa:2003qv} into the free quark Lagrangian. The Lagrangian density is written as~\cite{Klevansky:1992qe,Buballa:2003qv} 	
\begin{eqnarray}
	\mathcal{L}_{\rm NJL} &=& \bar{\psi}(i\gamma^\mu \partial_\mu -{\bf m})\psi 
+ G\sum_{a=0}^{8}\bigg[(\bar{\psi}\tau_{a}\psi)^2 + (\bar{\psi}i\gamma^5\tau_{a}\psi)^2\bigg]\nonumber\\
	&-&K \bigg[{\rm det}_{f}(\bar{\psi} (1+\gamma_{5})\psi + {\rm det}_{f}(\bar{\psi} (1-\gamma_{5})\psi \bigg]\label{AD1}
\end{eqnarray}
where $\mathbf{m}\equiv\begin{pmatrix}
	m_{u} & 0 & 0\\
	0 & m_{d} & 0\\
	0 & 0 & m_{s}
\end{pmatrix}$ and $\psi\equiv\begin{pmatrix}
	\psi_{u} \\
	\psi_{d}\\
	\psi_{s}
\end{pmatrix}$ are, respectively, the current quark mass matrix for u, d and s flavors and the field. In the present work, we assume the same mass for u and d quark flavors, i.e., $m_{u}=m_{d}$. $G$ is the coupling constant governing the scalar and pseudoscalar (four-fermion) interaction channels, $\tau_{a}(a=1,\dots, 8)$ represents the Gell-Mann matrices acting in the flavor space and $\tau_{0}=\frac{2}{3}\mathbb{1}_{f}$.
$K$ is the coupling constant corresponding to the six-fermion ’t Hooft interaction where the det$_{f}$ stands for the determinant in the flavor space. 
 The four and six-fermion contact interactions lead to quark self-energy, and one expresses the constituent quark mass $M_{i}$ in vacuum as~\cite{Klevansky:1992qe}, 
 \begin{eqnarray}
 	M_{i} &=&m_{i}+4iG\int_{\Lambda} \frac{d^{4}p}{(2\pi)^{4}}~{\rm Tr}S_{i}(p)-2K\bigg[\int_{\Lambda} \frac{d^{4}p}{(2\pi)^{4}}~{\rm Tr}S_{j}(p)\bigg]\nonumber\\  &&\bigg[\int_{\Lambda}\frac{d^{4}p}{(2\pi)^{4}}~{\rm Tr}S_{k}(p)\bigg],\text{where~} i\neq j\neq k~.\label{AD2}
 \end{eqnarray}
We have $S_{f}(p)=\frac{1}{\slashed{p}-M_{f}+i\epsilon}$ is the dressed quark propagator for the flavor $f$, the trace in the above equation is taken over color, and Dirac spaces. One may also express the gap-equation \eqref{AD2} in vacuum in terms of the quark condensate $\langle \bar{q}_{i}q_{i}\rangle$ as,
\begin{eqnarray}
	M_{i} &=&m_{i}-4G \langle \bar{q}_{i}q_{i} \rangle + 2K \langle \bar{q}_{j}q_{j}\rangle \langle \bar{q}_{k}q_{k}\rangle~,\text{where~} i\neq j\neq k\nonumber\\
	&=&m_{i}-4G\phi_{i}+2K\phi_{j}\phi_{k}\label{AD3}~.
\end{eqnarray}
The quark condensate is defined as~\cite{Klevansky:1992qe}
\begin{equation}
 \phi_{i}\equiv \langle \bar{q}_{i}q_{i}\rangle=-i\int_{\Lambda} \frac{d^{4}p}{(2\pi)^{4}}~{\rm Tr}S_{i}(p).
\end{equation}
 The NJL Lagrangian, containing the contact interaction terms, is non-renormalizable~\cite{Klevansky:1992qe}. Therefore, a cutoff energy scale $\Lambda$ is typically introduced to regularize the divergent integrals that appear in the calculations. In this work we use a sharp three-momentum cut-off $\Lambda$. The gap equation~\eqref{AD2} or \eqref{AD3} represents spontaneous chiral symmetry breaking in the vacuum, where a current quark of flavor $f$ with a light mass $m_{f}$ turns into a massive constituent quark with mass $M_{f}$. The parameters of the model---the current quark mass $m_f$, the coupling constants $G$ and $K$, and the cutoff scale $\Lambda$---are suitably chosen to reproduce the experimental values of the pion, kaon and eta prime meson masses to the values, $m_{\pi} = 135~\text{MeV}$, $m_{K}=497.7$ MeV, $m_{\eta^{\prime}}=957.8$ MeV and the pion decay constant, $f_{\pi} = 92.4~\text{MeV}$, in vacuum. In the present work, we use the following parameter values: $m_u=m_d= 5.5~\text{MeV}$, $m_s =140.7~\text{MeV}$, $\Lambda = 602.3~\text{MeV}$, $G\Lambda^2 = 1.835$, and $K\Lambda^5 =12.36$~\cite{Rehberg:1995kh}. In the presence of a medium at finite chemical potential $\mu_{f}$ and temperature $T$, the modified gap equation can be obtained by replacing the vacuum dressed propagator in \eqref{AD2} with the in-medium dressed propagator~\cite{Asakawa:1989bq,Klevansky:1992qe},
\begin{eqnarray}
	S_{i}(p,\mu_{i},T)&=&(\slashed{p}+M_{i})\bigg[\frac{1}{p^{2}-M_{i}^{2}+i\epsilon}+2\pi i~\delta(p^{2}-M_{i}^{2})\nonumber\\
	&&\left(\theta(p_{0})f^{0}+\theta(-p_{0})\bar{f}^{0}\right)\bigg]\nonumber,
\end{eqnarray}  
where the thermal distribution functions for quarks and anti-quarks are, respectively, $f^{0}=1/[e^{(E_{i}-\mu_{i})/T}+1]$, $\bar{f}^{0}=1/[e^{(E_{i}+\mu_{i})/T}+1]$ with $E_{i}=\sqrt{\vec{p}^{2}+M_{i}^{2}}$. Substituting the above propagator, the gap equation \eqref{AD2} after some simplification reads~\cite{Buballa:2003qv} ,
\begin{eqnarray}
	M_i&=&m_{i}+ 8GN_c\int_{\Lambda}\frac{d^{3}\vec{p}}{(2\pi)^{3}}\frac{M_{i}}{E_{i}}~(1-f_{i}^{0}-\bar{f}_{i}^{0})\nonumber\\
	&+&8KN_c^{2} \bigg[\int_{\Lambda}\frac{d^{3}\vec{p}}{(2\pi)^{3}}\frac{M_{j}}{E_{j}}~(1-f_{j}^{0}-\bar{f}_{j}^{0})\bigg]\nonumber\\
	&&\bigg[\int_{\Lambda}\frac{d^{3}\vec{p}}{(2\pi)^{3}}\frac{M_{k}}{E_{k}}~(1-f_{k}^{0}-\bar{f}_{k}^{0})\bigg] \label{AD4}
\end{eqnarray} 
where, $N_c=3$ is the color degeneracy. The quark condensates are obtained similarly as~\cite{Buballa:2003qv}  
\begin{eqnarray}
 \phi_{i}\equiv\langle \bar{q}_{i}q_{i}\rangle=-2N_{c}\int_{\Lambda} \frac{d^{3}\vec{p}}{(2\pi)^{3}} \frac{M_{i}}{E_{i}}~(1-f_{i}^{0}-\bar{f}_{i}^{0})~.\label{5AD}
\end{eqnarray}
The gap equations provided in Eq.~\eqref{AD4} can be also derived by minimizing the grand canonical potential (free energy) of the NJL Lagrangian. We pursue this approach briefly in this section. The NJL Lagrangian in the mean field approximation (MFA) can be expressed as~\cite{Ali:2024nrz,Gastineau:2001zke},
\begin{eqnarray}
\mathcal{L}_{\rm NJL}^{\rm MFA} &=& \bar{\psi}(i\gamma^\mu \partial_\mu -{\bf M})\psi -2G\sum_{f} \phi_{f}^{2}+4K\prod_{f} \phi_{f} \label{6AD}
\end{eqnarray}
where $\mathbf{M}\equiv\begin{pmatrix}
	M_{u} & 0 & 0\\
	0 & M_{d} & 0\\
	0 & 0 & M_{s}
\end{pmatrix}$ and $M_{f}$ is defined by Eq.~\eqref{AD3}. In this approach the value of the condensates $\phi_{f}$ are apriori unknown and has to be determined by minimizing the free energy $\Omega_{\rm NJL}$ of the system. The free energy from the mean field Lagrangian \eqref{6AD} is given by~\cite{Ali:2024nrz,Gastineau:2001zke}  
\begin{eqnarray}
	\Omega_{\rm NJL}&=&\Omega_{\rm MF} + \Omega_{\rm Vac} + \Omega_{\rm Th}, \text{ where}\label{7AD}\\
	\Omega_{\rm MF}&=& 2G\sum_{f} \phi_{f}^{2}-4K\prod_{f} \phi_{f},\label{8AD}\\
	 \Omega_{\rm Vac}&=& -2N_{c}\sum_{i}\int_{\Lambda}\frac{d^{3}\vec{p}}{(2\pi)^{3}} E_{i}, \label{9AD}\\
	 \Omega_{\rm Th}&=& -2N_{c}\sum_{i} \int\frac{d^{3}\vec{p}}{(2\pi)^{3}} \bigg[\ln(1+e^{-(E_{i}+\mu_{i})/T})\nonumber\\
	 &+&\ln(1+e^{-(E_{i}-\mu_{i})/T})\bigg]~.\label{10AD} 
\end{eqnarray}
The three gap equations provided in Eq.~\eqref{AD4} can be equivalently obtained from minimizing the free energy,
\begin{eqnarray}
 \frac{\partial \Omega_{\rm NJL} }{\partial \phi_{u}}= \frac{\partial \Omega_{\rm NJL} }{\partial \phi_{d}}= \frac{\partial \Omega_{\rm NJL} }{\partial \phi_{s}}=0~.\label{11AD}
\end{eqnarray}
In the present paper we only assume a baryon chemical potential, therefore, $\mu_{q}\equiv\mu_{u}=\mu_{d}=\mu_{s}=\frac{\mu_{B}}{3}$. Once the gap equations are solved to obtain the $M_{u} (\mu_{q},T)=M_{d} (\mu_{q},T)$ and $M_{s}(\mu_{q},T)$, one can determine the free energy $\Omega_{\rm NJL}(\mu_{B},T)$ from Eq.~\eqref{7AD}. The thermodynamic quantities of the system are determined in the usual way from the $\Omega_{\rm NJL}(\mu_{q},T)$.
\subsection{Chiral effective model}\label{chiral}
The chiral effective model has been used to study in-medium properties of finite nuclei, nuclear matter, strange baryonic matter \cite{Papazoglou:1997uw, Papazoglou:1998vr,Zschiesche:2002zr,Mishra:2003tr, Zschiesche:2003qq, Mishra:2006wy}. We use the mean-field chiral Lagrangian, based on chiral symmetry breaking and broken scale invariance, to obtain the expectation values of scalar and vector meson fields. The effective Lagrangian $\mathscr{L}$ in the mean-field approximation has the form,
\begin{equation}\label{L_density}
	\mathscr{L} = \mathscr{L}_{kin}+ \mathscr{L}_{BW}+ \mathscr{L}_{vec} +\mathscr{L}_{SSB} + \mathscr{L}_{scale-break} + \mathscr{L}_{ESB}~. 
\end{equation}
where, $\mathscr{L}_{kin}$ is the kinetic term,  $\mathscr{L}_{BW}$ is the baryon - meson interaction term where only scalar and vector mesons contribute, $\mathscr{L}_{vec}$ generates the mass of the vector mesons, $\mathscr{L}_{SSB}$ is the spontaneous chiral symmetry breaking term, $\mathscr{L}_{scale-break}$ is the QCD scale invariance breaking term, and $\mathscr{L}_{ESB}$ is the explicit chiral symmetry breaking term written in terms of the baryon field $\psi_i$, the  scalar-isoscalar meson fields $\sigma$ and $\zeta$, scalar-isovector meson field $\delta$, scalar-isoscalar dilaton field $\chi$ and vector fields $\rho$, $\omega$ and $\phi$,

\begin{eqnarray}
	&\mathscr{L}_{BW} = \mathscr{L}_{BX}+\mathscr{L}_{BV}\nonumber\\
	&= -\sum_{i=n,p} \bar{\psi}_i \left[m_i^*
	+ g_{i\omega}\gamma_0 \omega
	+ g_{i\rho}\gamma_0 \rho
	+ g_{i\phi}\gamma_0 \phi\right]\psi_i,\\[4pt]
	&\mathscr{L}_{vec} =
	\frac{1}{2}\left(\frac{\chi}{\chi_0}\right)^2
	\left(m_\omega^2\omega^2+m_\rho^2\rho^2+m_\phi^2\phi^2\right)\nonumber\\
	&+ g_4\left(\omega^4+6\rho^2\omega^2+\rho^4+2\phi^4\right),\\[4pt]
	&\mathscr{L}_{SSB} =
	-\frac{1}{2}k_0\chi^2(\sigma^2 + \zeta^2 + \delta^2)
	+ k_1(\sigma^2 + \zeta^2 + \delta^2)^2  \nonumber\\
	&\quad
	+ k_2\Bigg(\frac{\sigma^4}{2} + \frac{\delta^4}{2}
	+ 3\sigma^2\delta^2 + \zeta^4\Bigg)
	+ k_3\chi(\sigma^2 - \delta^2)\zeta
	- k_4\chi^4,\\[4pt]
	&\mathscr{L}_{scale\text{-}break} =
	- \frac{1}{4}\chi^4 \ln{\frac{\chi^4}{\chi_0^4}}
	+ \frac{d}{3}\chi^4 \ln{\Bigg(
		\frac{(\sigma^2 - \delta^2)\zeta}{\sigma_0^2\zeta_0}
		\Big(\frac{\chi}{\chi_0}\Big)^3
		\Bigg)}, \label{sb}\\[4pt]
	&\mathscr{L}_{ESB} =
	-\Big(\frac{\chi}{\chi_0}\Big)^2
	\mathrm{Tr}\Bigg[\mathrm{diag}\Bigg(
	\frac{1}{2}m_\pi^2f_\pi(\sigma + \delta),~
	\frac{1}{2}m_\pi^2f_\pi(\sigma - \delta),~ \nonumber\\
	&\qquad
	\Big(\sqrt{2}m_K^2f_K
	- \frac{1}{\sqrt{2}}m_\pi^2f_\pi \Big)\zeta
	\Bigg)\Bigg].
	\label{L_ESB}
\end{eqnarray}
where we consider nuclear matter with protons and neutrons $(i=p,n)$ with effective in-medium nucleon mass  $m^*_i = -g_{\sigma i} \sigma - g_{\zeta i}\zeta - g_{\delta i}\delta$. 
The equations of motion for the scalar and vector fields then have the form,

\begin{eqnarray}
	\label{l1}
	&k_0\chi^2\sigma-4k_1\sigma(\sigma^2+\zeta^2+\delta^2)-2k_2(\sigma^3+3\sigma\delta^2)-2k_3\chi\sigma\zeta \nonumber\\
	&-\frac{d}{3}\chi^4\Big(\frac{2\sigma}{\sigma^2-\delta^2}\Big)+\Big(\frac{\chi}{\chi_0}\Big)^2m_\pi^2f_\pi-\sum_i g_{\sigma i}\rho_i^s =0,\\
	\label{l2}
	&k_0\chi^2\zeta-4k_1\zeta(\sigma^2+\zeta^2+\delta^2)-4k_2\zeta^3-k_3\chi(\sigma^2-\delta^2)-\frac{d}{3}\frac{\chi^4}{\zeta}\nonumber\\
	&+\Big(\frac{\chi}{\chi_0}\Big)^2\Bigg[\sqrt{2}m_K^2f_K-\frac{1}{\sqrt{2}}m_\pi^2f_\pi\Bigg]-\sum_i g_{\zeta i}\rho_i^s=0,\\
	\label{l3}
	&k_0\chi^2\delta-4k_1\delta(\sigma^2+\zeta^2+\delta^2)-2k_2\delta(\delta^2+3\sigma^2) 
	+2k_3\chi\delta\zeta \nonumber\\
	&+\frac{2}{3}d\chi^4\Bigg(\frac{\delta}{\sigma^2-\delta^2}\Bigg)-\sum_i g_{\delta i}\rho_i^s=0,\\
	\label{l4}
	&k_0\chi(\sigma^2+\zeta^2+\delta^2)-k_3\zeta(\sigma^2-\delta^2)+\chi^3\Bigg[1+4\ln{\frac{\chi}{\chi_0}}\Bigg] \nonumber\\
	&+(4k_4-d)\chi^3-\frac{4}{3}d\chi^3\ln{\Bigg[\Big(\frac{(\sigma^2-\delta^2)\zeta}{\sigma_0^2\zeta_0}\Big)\Big(\frac{\chi}{\chi_0}\Big)^3\Bigg]} \nonumber\\
	&+\Big(\frac{2\chi}{\chi_0^2}\Big)\Bigg[m_\pi^2f_\pi\sigma+\Bigg(\sqrt{2}m_K^2f_K-\frac{1}{\sqrt{2}}m_\pi^2f_\pi\Bigg)\zeta\Bigg] = 0,\\
	&\left(\frac{\chi}{\chi_0}\right)^{2} m_{\omega}^{2}\,\omega
	+ 4g_{4}\left(\omega^{3} + 3\rho^{2}\omega\right)
	- \sum_{i} g_{\omega i}\,\rho_{i} = 0,\\
	&\left(\frac{\chi}{\chi_0}\right)^{2} m_{\rho}^{2}\,\rho
	+ 4g_{4}\left(\omega^{3} + 3\rho^{2}\omega\right)
	- \sum_{i} g_{\rho i}\,\rho_{i} = 0,\\
	&\left(\frac{\chi}{\chi_0}\right)^{2} m_{\phi}^{2}\,\phi
	+ 8g_{4}\,\phi^{3}
	- \sum_{i} g_{\phi i}\,\rho_{i} = 0.
\end{eqnarray}
Here, the net baryon density $\rho_i\equiv\langle \psi_{i}^\dagger\psi_{i}\rangle$ and the net scalar density $\rho_i^s\equiv\langle \bar{\psi_{i}} \psi_{i}\rangle$ are defined as
\begin{eqnarray}
	&\rho_i =2\int\frac{d^3p}{(2\pi)^3}~ (f^{*}_{i}-\bar{f}^{*}_{i})\\
	&\rho_i^s =2\int\frac{d^3p}{(2\pi)^3}~\frac{m_i^*}{E_i^*} (f^{*}_{i}+\bar{f}^{*}_{i}),
\end{eqnarray}
where the proton (neutron) distribution is \\$f^{*}_{i}=1/[e^{(E^{*}_{i}-\mu_{i}^*)/T}+1]$ and anti-proton (anti-neutron) distribution is $\bar{f}^{*}_{i}=1/[e^{(E^{*}_{i}+\mu_{i}^*)/T}+1]$ with $E_{i}^{*}=\sqrt{p^{2}+m_{i}^{*2}}$ and $\mu_i^*=\mu_i-(g_{\rho i}\tau_3 \rho+g_{\omega i}\omega+g_{\phi i}\phi)$. 
Solving the equations of motion for the scalar and vector fields, we obtain their values as a function of baryon density $\rho_{B}$ and temperature $T$. 

The trace of the energy-momentum tensor in QCD has the form,
\begin{eqnarray}
	T_\mu^\mu=\left\langle\frac{\beta_{QCD}}{2g}G^a_{\mu\nu}G^{\mu\nu a}\right\rangle+\sum_{i=u,d,s}m_i\bar{q}_iq_i
\end{eqnarray}
where the first term corresponds to the gluon condensate (accounted by the dilaton field in the effective theory) and the second term generates the finite quark masses and can be identified as the negative of the explicit chiral
symmetry-breaking term in the chiral Lagrangian.
For isospin symmetric matter with $\delta=0$ we obtain the non-strange and strange quark condensates as $\langle \bar{u}u \rangle = \frac{1}{2m_u} m_\pi^2f_\pi\sigma$, $\langle \bar{d}d \rangle = \frac{1}{2m_d} m_\pi^2f_\pi\sigma$ and $\langle \bar{s}s \rangle=\frac{1}{m_s}\left(\sqrt{2}m_K^2f_K - \frac{1}{\sqrt{2}}m_\pi^2f_\pi\right) \zeta$ in the frozen glue-ball limit $\chi=\chi_0$.We take the values of the current quark masses to be  $m_u=m_d=4.6$ MeV and $m_s=95$ MeV.

Using the density dependent values of $\sigma$ and $\zeta$ in the expression of $\langle \bar{q_i}q_i\rangle$ where $q_i=\{u,d,s\}$ one can obtain the density-dependent quark condensates $\langle \bar{u}u\rangle$, $\langle \bar{d}d\rangle$ and $\langle \bar{s}s \rangle $. Motivated by NJL-type models~\cite{Buballa:2003qv,Buballa:2014tba}, normalizing the density-dependent quark condensate with its vacuum values, we express the constituent quark masses for all three light quark flavors as \cite{Marattukalam:2024mef},
\begin{align}
	M_{q_i}(n_B,T)=& ~m_{q_i}+\frac{\langle \bar{q_i}q_i\rangle_{(n_B,T)}}{\langle \bar{q_i}q_i\rangle_{(0)}}~(M_{q_i(0)}-m_{q_i})\label{M_q}~,
\end{align}
where $n_B$ is the net baryon density, $\langle \bar{q_i}q_i\rangle_{(0)}$ and  $M_{q_i(0)}$ are the vacuum expectation values of the quark condensates and the constituent quark masses respectively. The vacuum expectation value of the $u$ and $d$ constituent quark masses is fixed at one-third the mass of proton in vacuum, \textit{i.e.,} $M_{u(0)} =M_{d(0)} = 939/3 $~ MeV = 313 MeV and that of the $s$ quark is fixed at $M_{s(0)} =489$~MeV to fit the mass of $\Lambda^0$ baryon with vacuum mass 1115 MeV which contains one each of the $u$, $d$ and $s$ quarks. Reader may note that this prescription for constituent quark mass was adapted to enable the use of a quasiparticle picture of quarks and is different from the original hadronic chiral effective model.

As stated earlier, in the present work we consider only a baryon chemical potential, $\mu_{q}\equiv\mu_{u}=\mu_{d}=\mu_{s}=\frac{\mu_{B}}{3}$. 
\subsection{Hadron Resonance Gas (HRG) model}\label{HRG}
The HRG model describes the hadronic medium as a non-interacting gas of all known hadrons and resonances. The HRG formalism treats the confined phase of QCD as a gas of point-like hadrons and resonances, which serve as the effective degrees of freedom. In our analysis, we include all hadronic states with masses up to 2.6 GeV~\cite{ParticleDataGroup:2020ssz}. The thermodynamic and transport properties of the hadronic medium can thus be written as the sum over individual particles. The grand canonical partition function for a single hadron species $i$ is given by
\begin{equation}
	\ln Z_i(T, V, \mu_i) = \pm V g_i \int \frac{d^3p}{(2\pi)^3} \ln\left[1 \pm \exp\left(-\frac{E_{i} - \mu_i}{T}\right)\right]\label{HRG1}
\end{equation}
with $(+)$ for fermions (baryons)  
and $(-)$ for bosons (mesons) and $g_i=2s_i+1$, $E_{i}=\sqrt{p^{2}+m_{i}^{2}}$ is the spin degeneracy and energy for the $i^\text{th}$ species respectively. $ \mu_i = B_i \mu_B$ is the baryon chemical potential for species $ i $, determined by its baryon number $B_i$ (we assume the electric charge and strange chemical potentials to be zero). The total partition function of the HRG is obtained by summing over all hadron species,
\begin{equation}
	\ln Z_{\text{HRG}} = \sum_i \ln Z_i ~. \label{HRGgrand}
\end{equation}
The usual relation gives the thermodynamic quantities,
\begin{eqnarray}
	&&  P=-\Omega,~n_{B}=-\bigg(\frac{\partial \Omega}{\partial \mu_{B}}\bigg)_{T}~,\nonumber\\
	&&s=-\bigg(\frac{\partial \Omega}{\partial T} \bigg)_{\mu_{B}},~ \varepsilon=Ts+\mu_{B}n_{B}-P~,\label{HRG2}
\end{eqnarray}
where the grand canonical potential $\Omega=-\frac{T}{V}\ln Z_{\rm HRG}$. Using equation \eqref{HRGgrand} in equation \eqref{HRG2}, we can get all thermodynamical quantities in terms of thermal distribution functions of different hadrons. The total energy density, pressure, net baryon density, total number density of hadrons, and total entropy density are given by,
\begin{eqnarray}
	&&\varepsilon = 
	\sum_{M}g_{M}\int \frac{d^3p}{(2\pi)^3}\, E_M \, f_M^0
	+\sum_{B}g_{B}\int \frac{d^3p}{(2\pi)^3}\, E_B \,f_B^0~,
	\label{totalenergy}\\
	&&P =
	\sum_{M} g_{M}\int \frac{d^3p}{(2\pi)^3}\, \frac{p^2}{3E_M} \, f_M^0
	+\sum_{B} g_{B}\int \frac{d^3p}{(2\pi)^3}\, \frac{p^2}{3E_B} \, f_B^0 ,\nonumber\\
	\label{totalpressure}\\
	&&n_B =
	\sum_{B}B_{Q}g_{B}\int \frac{d^3p}{(2\pi)^3}\,
	f_B^0,
	\label{netbaryon}\\
	&&n =
	\sum_{M}g_{M}\int \frac{d^3p}{(2\pi)^3}\, f_M^0
	+\sum_{B}g_{B}\int \frac{d^3p}{(2\pi)^3}\,
	f_B^0
	\label{totalhadron}\\
	&&s =
	\frac{\varepsilon + P - \mu_B n_B}{T}.
	\label{entropydensity}
\end{eqnarray}
where $f_M^0=1/[e^{(E_M/T)}-1]$ is Bose-Einstein distribution function for meson with energy $E_M=\sqrt{p^2+m_M^2}$, $f_B^0=1/[e^{(E_B-B_{Q}\mu_B)/T}+1]$ are Fermi-Dirac distribution function for baryons having energy $E_B=\sqrt{p^2+m_B^2}$ and baryon number $B_{Q}$ ($+1$ for baryons and $-1$ for anti-baryons), $\sum_{M}$ and $\sum_{B}$ stand for summation over all mesons and baryons. Above thermodynamical expressions for HRG model are valid within the temperature and net baryon chemical potential domain, where hadrons are constituent particles of the medium.

For quark degrees of freedom, the net quark density $n_q$ is related to the net baryon density $n_{B}$ as $n_q=3n_{B}$,
\begin{align}
	&n_{q}=3n_{B}= 2\sum_{i=u,d,s}N_c\int \frac{d^3\vec{p}}{(2\pi)^3}~ [f_i^0(T,\mu_q)-\bar{f}^0_i(T,\mu_q)] ,\label{QFTnq}
\end{align}
where $f^{0}_i=1/[e^{(E_{i}-\mu_{q})/T}+1]$ and $\bar{f}^{0}_q=1/[e^{(E_{i}+\mu_{q})/T}+1]$ stands for quark and anti-quark distributions for the $i$ th flavor respectively with the modified dispersion relation $E_{i}=\sqrt{p^{2}+M_{i}^{2}}$. The constituent mass of quark flavors $M_{f}$ for NJL and chiral effective models are different due to their different background frameworks.
For the NJL model net baryon density $n_{B}(\mu_{q},T)$ can be obtained from Eq.~\eqref{QFTnq} with the use of constituent quark mass $M_{f}=M_{f}(\mu_{q},T)$ obtained by solving Eq.\eqref{AD4}. Whereas for the chiral effective model, the expression for the quark chemical potential $\mu_q(n_{B},T)$ can be obtained self-consistently from Eq.~\eqref{QFTnq} by using the constituent quark mass $M_{f}(n_{B},T)$ provided in Eq.~\eqref{M_q}. By putting $M_f=0$, one can obtain massless quark matter results, which will be discussed in our results section. Other thermodynamical quantities for quark matter are 
\begin{eqnarray}
	&&P=2N_c \sum_{i}\int \frac{d^3\vec{p}}{(2\pi)^3}\frac{p^2}{3E} (f_i^0+\bar{f}_i^0),\label{njlthp}\\
	&&\varepsilon=2N_c\sum_{i} \int \frac{d^3\vec{p}}{(2\pi)^3}E (f_i^0+\bar{f}_i^0),\label{njlthe}\\
	&&s=\frac{\varepsilon +P -\mu_q n_q}{T}\label{njlths}~.
\end{eqnarray}
We note that for the NJL model the pressure and energy density has contribution from vacuum as well as the mean field parts, i.e., $P_{\rm NJL}=-\Omega_{\rm NJL}=P_{\rm Th}+P_{\rm MF}+P_{\rm Vac}= P_{\rm Th}-B$, where $P_{\rm MF}=-\Omega_{\rm MF}$, $P_{\rm Vac}=-\Omega_{\rm Vac}$ and $B\equiv\Omega_{\rm MF}+\Omega_{\rm Vac}$. Similarly the energy density $\varepsilon_{\rm NJL}=\varepsilon_{\rm Th}+B$ such that $P_{\rm NJL}+\varepsilon_{\rm NJL}=P_{\rm Th}+\varepsilon_{\rm Th}$. Only the thermal parts of the pressure and energy density are defined by Eqs.~\eqref{njlthp} and~\eqref{njlthe}. The entropy density and net quark density receives no contribution from the mean fields or vacuum which can be checked explicitly by differentiating the $\Omega_{\rm NJL}$ with respect to $\mu_{q}$ or $T$. In the Sec.~\eqref{sec:results} while presenting results of the thermodynamic quantities in NJL model the total pressure and total energy density (including the mean field and vacuum parts) is displayed.
The total quark number density is given by,
\begin{eqnarray}
	&&  n=2N_c \sum_{i} \int \frac{d^3\vec{p}}{(2\pi)^3} (f_i^0+\bar{f}_i^0).
\end{eqnarray}

After discussing the three different models used to effectively describe the thermodynamics of QCD matter at finite temperature and density, we now turn our attention to hydrodynamics (next subsection), which provides a suitable framework to describe the evolution of QCD matter produced in heavy-ion collisions. 
\subsection{Transport coefficients}
\label{thermo-transport}
The nuclear matter produced in heavy-ion collisions undergoes expansion and cooling. During this evolution, it can be effectively described by hydrodynamics — a theory that captures the large-time and long-wavelength behavior of a non-equilibrium many-body system~\cite{Denicol2021ch1}. The macroscopic quantities of fundamental importance in hydrodynamic descriptions are the energy–momentum flow, $T^{\mu\nu}$, and the conserved four-currents associated with baryon number and electric charge, $J_{B}^{\mu}$ and $J_{Q}^{\mu}$, respectively. In the non-equilibrium scenario the flows can be described by splitting them into two parts--equilibrium parts: $T^{(0)\mu\nu}$, $J_{B}^{(0)\mu}$, and $J_{Q}^{(0)\mu}$ and non equilibrium parts: $\delta T^{\mu\nu}$, $\delta J_{B}^{\mu}$, and $\delta J_{Q}^{\mu}$. The division of the flows into equilibrium and non-equilibrium parts becomes manifest from the following kinetic theory definitions~\cite{DeGroot:1980dk}, 
\begin{eqnarray} 
	&&J_{B}^{\mu}=J_{B}^{(0)\mu}+\delta J_{B}^{\mu}= \sum_{k}g_{k}B_{k}\int\frac{d^3\vec{p}_{k}}{(2\pi)^3} \frac{p_{k}^{\mu}}{E_{k}} (f^{0}_{k}+\delta f_{k}),\nonumber\\
	&&J_{Q}^{\mu}=J_{Q}^{(0)\mu}+\delta J_{Q}^{\mu}= \sum_{k}g_{k}Q_{k}\int\frac{d^3\vec{p}_{k}}{(2\pi)^3} \frac{p_{k}^{\mu}}{E_{k}} (f^{0}_{k}+\delta f_{k}),\nonumber\\
	&&T^{\mu\nu}=T^{(0)\mu\nu}+\delta T^{\mu\nu}=\sum_{k}g_{k}\int\frac{d^3\vec{p}_{k}}{(2\pi)^3}\frac{p_{k}^{\mu}p_{k}^{\nu}}{E_{k}} (f^{0}_{k}+\delta f_{k}),\nonumber\\
	\label{Hydro1}
\end{eqnarray}
where the particle index $k=\{u,\bar{u},d,\bar{d},s,\bar{s}\}$ for a system described by NJL and chiral effective model whereas $k=\{\text{all hadrons}\}$ in the case of HRG. The particle distribution $f_{k}$ is written as the sum of equilibrium distribution $f^{0}_{k}=1/[e^{(E_{k}-\mu_{k})/T}-a_{k}]$ ($a_{k}=-1$ for fermions and $a_{k}=+1$ for bosons) and a perturbation  $\delta f_{k}$ , i.e., $f_{k}=f^{0}_{k}+\delta f_{k}$. The electric charge, baryon number, and spin degeneracy for the $k$ th species are, respectively, $Q_{k}$, $B_{k}$, and $g_{k}$. The chemical potential for the $k$ th species is given by $\mu_{k}=B_{k}\mu_{B}$. Mainly, the baryons have non-zero $B_k=\pm1$, and the mesons have $B_k=0$. For quark flavors or species, we have $\mu_k=\pm \frac{1}{3} \mu_B$ for quark/anti-quark. One normally defines $\mu_{q}=-\mu_{\bar{q}}=\frac{\mu_{B}}{3}$ for the quarks where net quark density $n_{q}=3n_{B}$. We assume the electric charge and strange chemical potentials to vanish on average, $\mu_{Q}=\mu_s=0$. The thermodynamic variables can be extracted from the equilibrium part of the flows as,
\begin{eqnarray} 
	&& n_{B}= u_{\mu}J_{B}^{(0)\mu}, n_{Q}=u_{\mu}J_{Q}^{(0)\mu}, \varepsilon=u_{\mu}u_{\nu}T^{(0)\mu\nu}.\label{Hydro2}
\end{eqnarray}
By going to the local rest frame (LRF) of the fluid, $u^{\mu} = (1, \vec{0})$, one can verify that the above definitions of the thermodynamic variables are consistent with those statistical mechanical base expression given in Secs.~\ref{NJL}, \ref{chiral}, and \ref{HRG}. Transport coefficients, such as the thermal conductivity $\kappa$, the electrical conductivity $\sigma$, and the shear viscosity $\eta$, play a role in the dissipative or non-equilibrium part of the flows defined in \eqref{Hydro1}. They can be determined by calculating the dissipative part of the flows microscopically and then comparing them with the usual macroscopic laws~\cite{DeGroot:1980dk},
\begin{eqnarray}
	&& \delta J_{q}^{i}\equiv\delta T^{0i}-h\delta J_{B}^{i}=-\kappa \partial_{i}T \text{ (Fourier's law)},\label{Hydro3}\\
	&& \delta J_{Q}^{i}=\sigma \tilde{E}^{i} \text{ (Ohm's law)},\label{Hydro4}\\
	&& \delta T^{ij}= -2\eta U^{ij} \text{ (Newton's law)},\label{Hydro5}
\end{eqnarray}
where $\tilde{E}^{i}$ is the applied electric field, $U^{ij}\equiv(\frac{1}{2}(\partial_{i}u^{j}+\partial_{j}u^{i})-\frac{1}{3}\delta^{ij}\partial_{l}u^{l})$ is the traceless fluid field gradient, and $h=\frac{\varepsilon+P}{n_{B}}$ is the enthalpy per baryon number. The microscopic evaluation of the dissipative flows can be performed by getting the form of $\delta f_{k}$ from the BTE and evaluating the integrals \eqref{Hydro1}. In the following, we briefly explain the steps to obtain $\delta f_{k}$ and the subsequent evaluation of the transport coefficients  $\kappa$, $\sigma$, and $\eta$. More details on the procedure can be found in \cite{DeGroot:1980dk,Cercignani2002,Dwibedi:2024mff,Denicol2021}. The linearized BTE for the $k$ th species in the relaxation time approximation (RTA) is given by,
\begin{eqnarray}
	&& \frac{\partial f^{0}_{k}}{\partial t}+\frac{p^{i}_{k}}{E_{k}}\frac{\partial f^{0}_{k}}{\partial x^{i}}+F_{k}^{i}\frac{\partial f^{0}_{k}}{\partial p^{i}_{k}}=-\frac{\delta f_{k}}{\tau_{c}},\label{Hydro6}
\end{eqnarray}
where $\tau_{c}$ is the relaxation time and $F^{i}_{k}$ is the external force on the $k$ th species. From Eq.\eqref{Hydro6}, we see for a steady and homogeneous system the out-of-equilibrium distribution $\delta f_{k}$ vanishes identically in the absence of external forces. The system can be driven out of equilibrium by introducing inhomogeneities in the spatio-temporal profiles of the temperature, $T(x^{i}, t)$, the baryon chemical potential, $\mu_{B}(x^{i}, t)$, or the fluid velocity, $u^{i}(x^{i}, t)$, or by applying an external force $F^{i}_{k}=Q_{k}\tilde{E}^{i}$. For the determination of thermal conductivity, let us drive the system out of equilibrium by creating a spatially inhomogeneous temperature $T(x^{i})$ and baryon chemical potential $\mu_{B}(x^{i})$ profile. In this scenario, the BTE in RTA can be used to get,
\begin{eqnarray}
	&&\delta f_{k}=\tau_{c}f^{0}_{k}(1+a_{k}f^{0}_{k})\left[E_{k}-hB_{k}\right]\frac{p^{j}_{k}}{E_{k}}\partial_{j}\frac{1}{T}~,\label{Hydro7}
\end{eqnarray} 
where we have used the Gibbs-Duhem relation \cite{Gavin:1985ph} to eliminate the spatial gradient of the baryon chemical potential. Substituting the above expression in the definition of the heat flow $\delta J_{q}^{i}\equiv \delta T^{0i}-h\delta J_{B}^{i}$, we have,
\begin{eqnarray}
	\delta J_{q}^{i}&=&\sum_{k}\int \frac{d^3\vec{p}_{k}}{(2\pi)^3} [E_{k}-hB_{k}]\frac{p^{i}_{k}}{E_{k}}\delta f_{k}\nonumber \\
	&=&- \frac{\tau_{c}}{3T^{2}}\sum_{k}g_{k}\int \frac{d^3 p_{k}}{(2\pi)^3} [E_{k}-hB_{k}]^{2}\frac{p^{2}_{k}}{E^{2}_{k}}\nonumber\\
	&&f^{0}_{k}(1+a_{k}f^{0}_{k})\partial_{i}T,\nonumber
\end{eqnarray}
comparing it with the macroscopic Fourier's law (Eq.~\eqref{Hydro3}), we have,
\begin{eqnarray}
	\kappa&=& \frac{\tau_{c}}{3T^{2}}\sum_{k}g_{k}\int \frac{d^3 p_{k}}{(2\pi)^3} [E_{k}-hB_{k}]^{2}\frac{p^{2}_{k}}{E^{2}_{k}}f^{0}_{k}(1+a_{k}f^{0}_{k})~.\nonumber\\\label{Hydro8}
\end{eqnarray}
Similarly, to obtain the electrical conductivity, one needs to apply only the external force $F^{i}_{k}=Q_{k}\tilde{E}^{i}$ to drive the system out of equilibrium. In this case, the BTE in RTA gives,
\begin{eqnarray}
	&&\delta f_{k}=\tau_{c}f^{0}_{k}(1+a_{k}f^{0}_{k})\frac{p^{j}_{k}}{E_{k}T}Q_{k}\tilde{E}^{i}~.\label{Hydro9}
\end{eqnarray}
Using this to evaluate $\delta J_{Q}^{i}$ and comparing the resulting expression with Eq.~\eqref{Hydro4} we have,
\begin{eqnarray}
	&&\sigma=\frac{\tau_{c}}{3T}\sum_{k}g_{k}Q_{k}^{2}\int \frac{d^3 p_{k}}{(2\pi)^3} \frac{p^{2}_{k}}{E^{2}_{k}}f^{0}_{k}(1+a_{k}f^{0}_{k})~.\label{Hydro10}
\end{eqnarray} 
The shear viscosity can be determined by creating an inhomogeneous fluid profile $u^{i}(x^{i})$ while keeping the other thermodynamic quantities homogeneous. Therefore, $\delta f_{k}$ for the determination of the shear viscous flows can be written as,
\begin{eqnarray}
	&&\delta f_{k}=-\tau_{c}f^{0}_{k}(1+a_{k}f^{0}_{k})\frac{p^{l}_{k}p^{m}_{k}}{E_{k}T}U^{lm}~,\label{Hydro11}
\end{eqnarray}
where in obtaining this, we ignore the bulk viscous effects. Substituting this in the evaluation of $\delta T^{ij}$ and comparing with the macroscopic law \eqref{Hydro5} we have,
\begin{eqnarray}
	&&\eta=\frac{\tau_{c}}{15T}\sum_{k}g_{k}\int \frac{d^3 p_{k}}{(2\pi)^3} \frac{p^{4}_{k}}{E^{2}_{k}}f^{0}_{k}(1+a_{k}f^{0}_{k})~.\label{Hydro10}
\end{eqnarray}

\section{Results and discussion}
\label{sec:results}

In this section, we present a quantitative analysis and comparison of thermodynamic and transport properties of hot and dense matter as described by various effective theories -- the NJL model, the chiral effective model, and the HRG model. The temperature dependence of these properties is well documented in the context of RHIC and LHC experiments. We systematically extend this study to the baryon-rich, dense sector in the context of upcoming CBM and NICA experiments, where a finite baryon density is expected.

\begin{figure}[htbp]
	\centering
	
	\begin{subfigure}{0.24\textwidth}
		\centering
		\includegraphics[width=\linewidth]{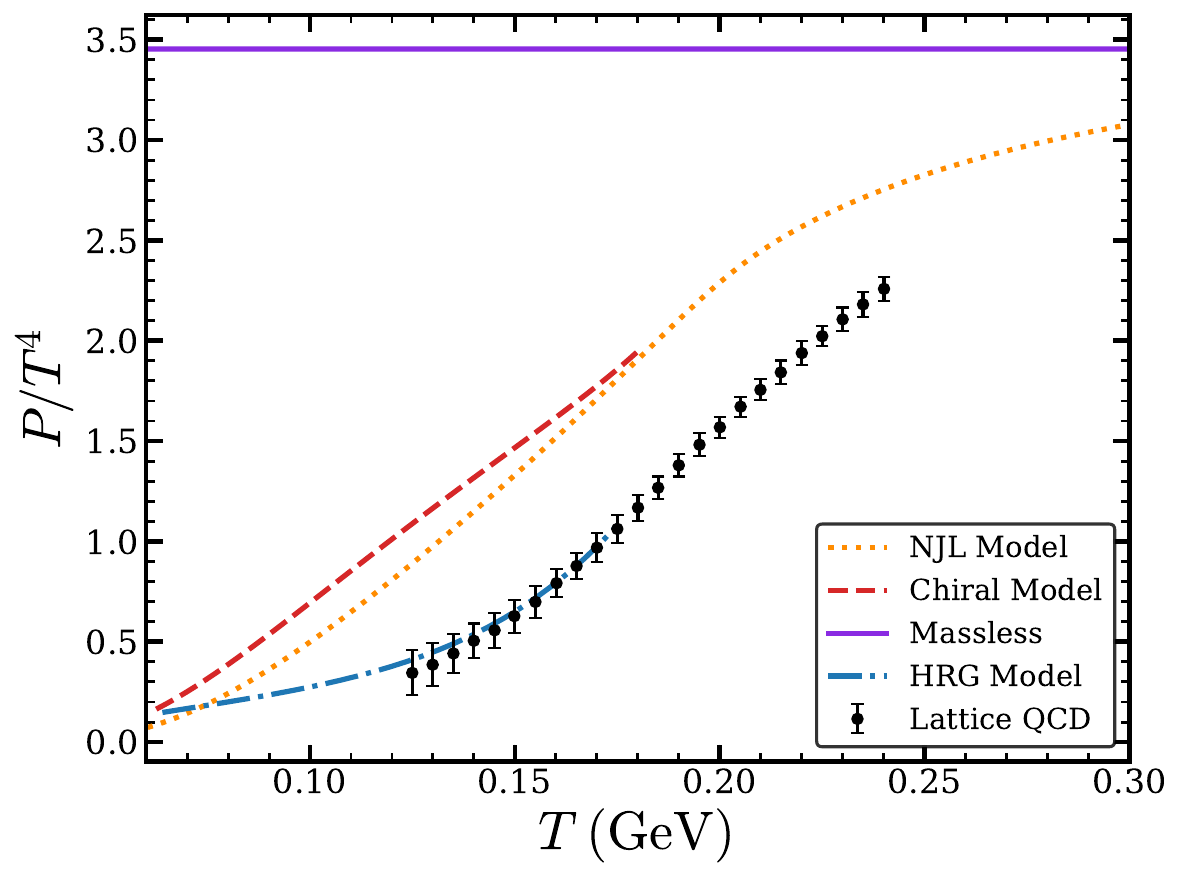}
	\end{subfigure}
	\hfill
	\begin{subfigure}{0.24\textwidth}
		\centering
		\includegraphics[width=\linewidth]{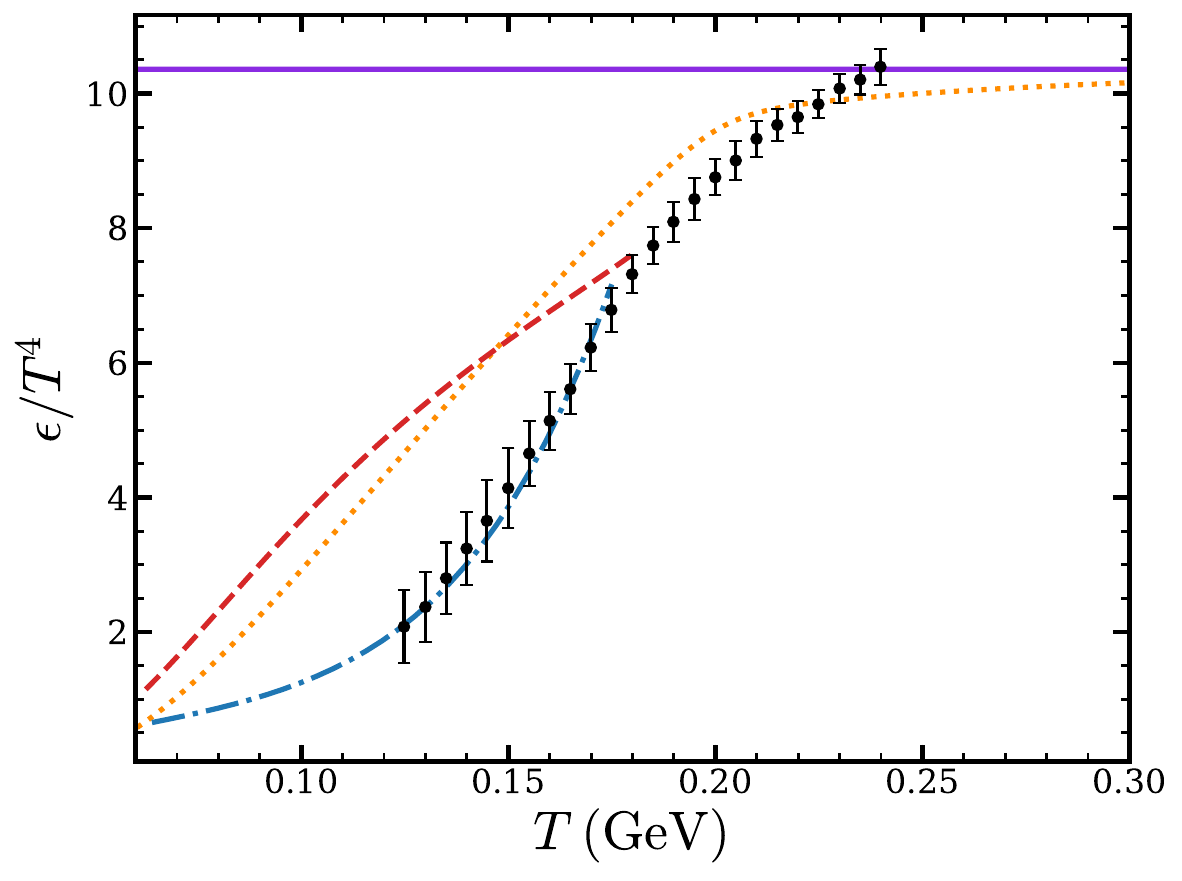}
	\end{subfigure}
	
	\vspace{0.5cm}
	
	\begin{subfigure}{0.24\textwidth}
		\centering
		\includegraphics[width=\linewidth]{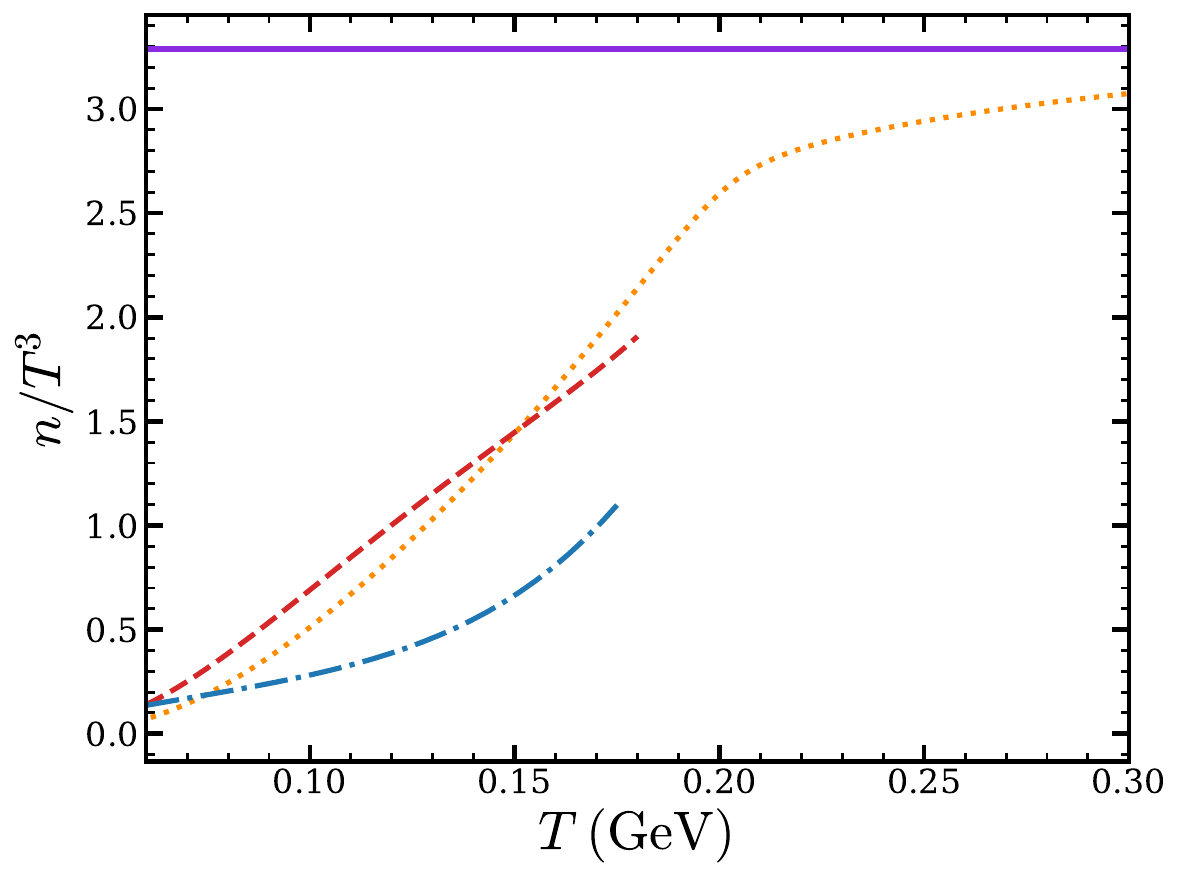}
	\end{subfigure}
	\hfill
	\begin{subfigure}{0.24\textwidth}
		\centering
		\includegraphics[width=\linewidth]{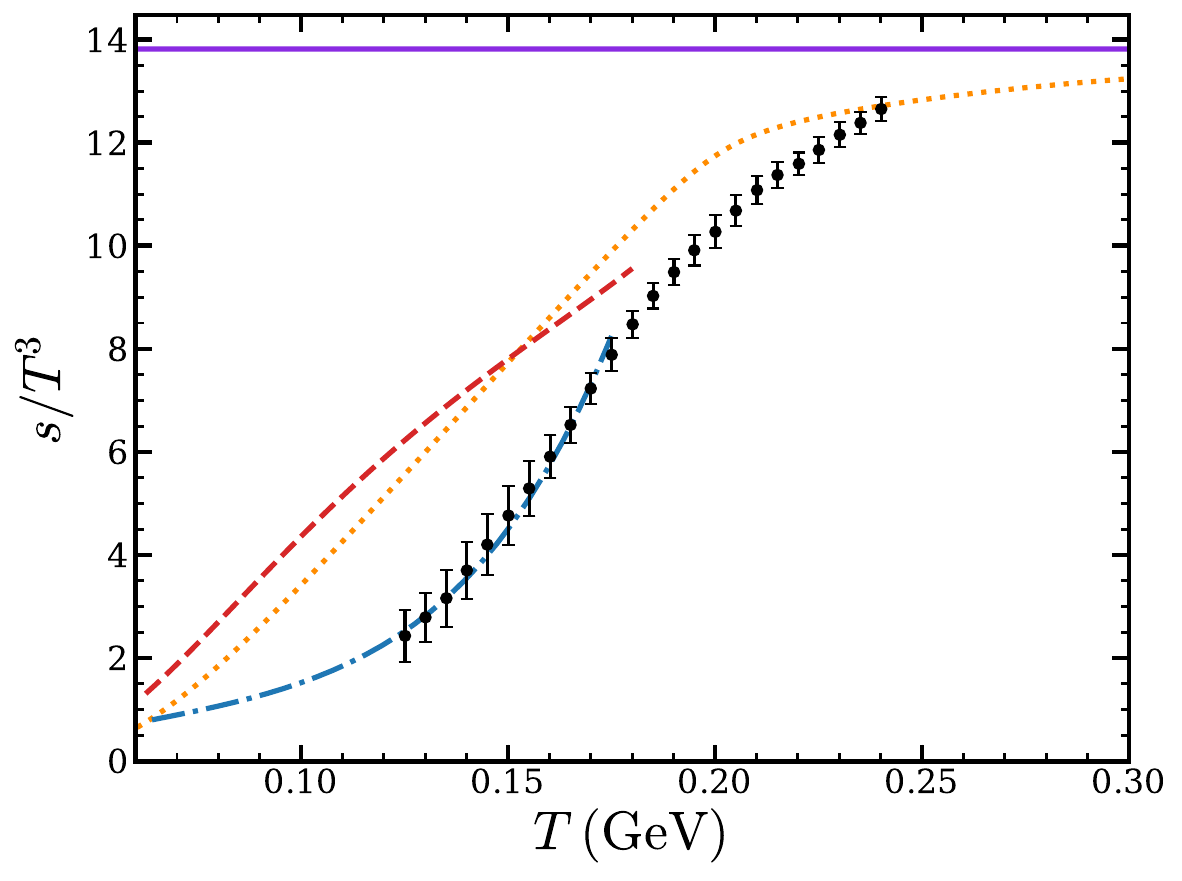}
	\end{subfigure}
	
	\vspace{0.5cm}
	
	\begin{subfigure}{0.24\textwidth}
		\centering
		\includegraphics[width=\linewidth]{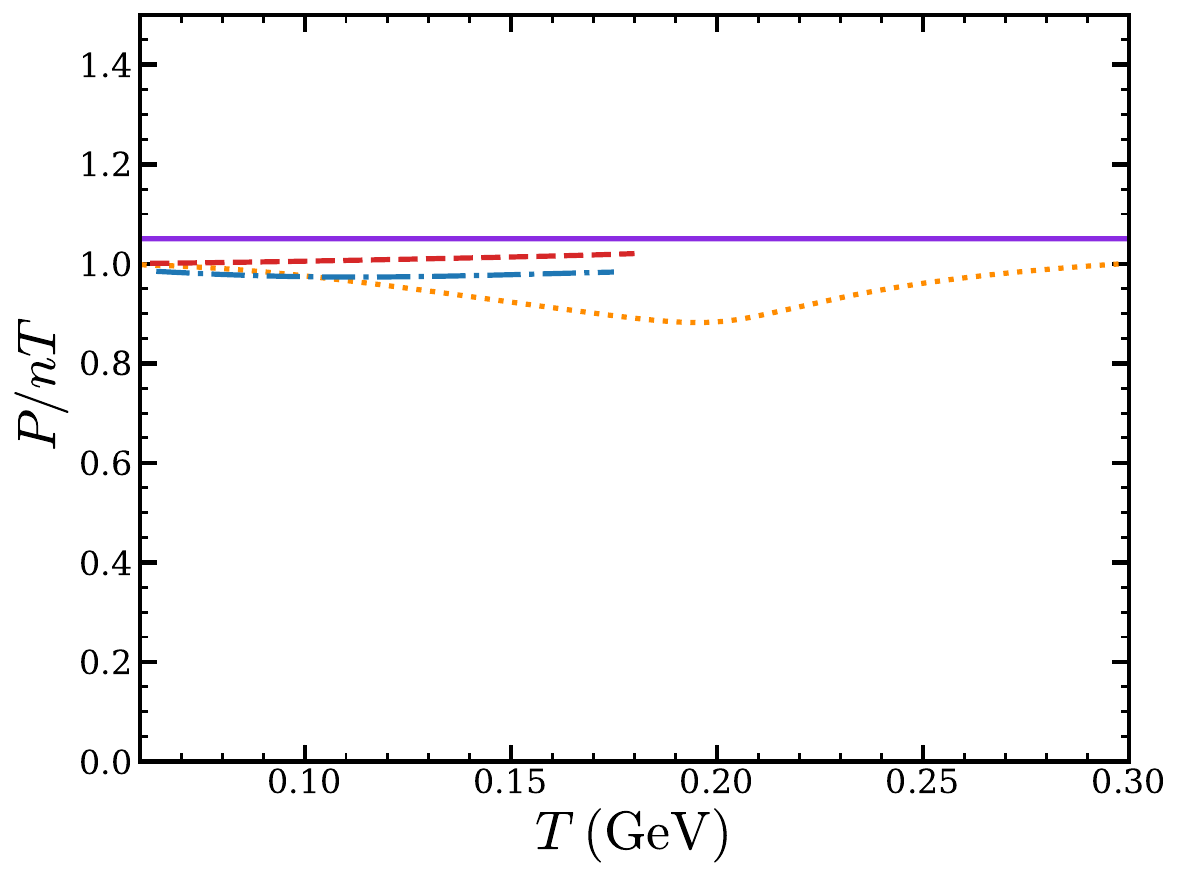}
	\end{subfigure}
	\hfill
	\begin{subfigure}{0.24\textwidth}
		\centering
		\includegraphics[width=\linewidth]{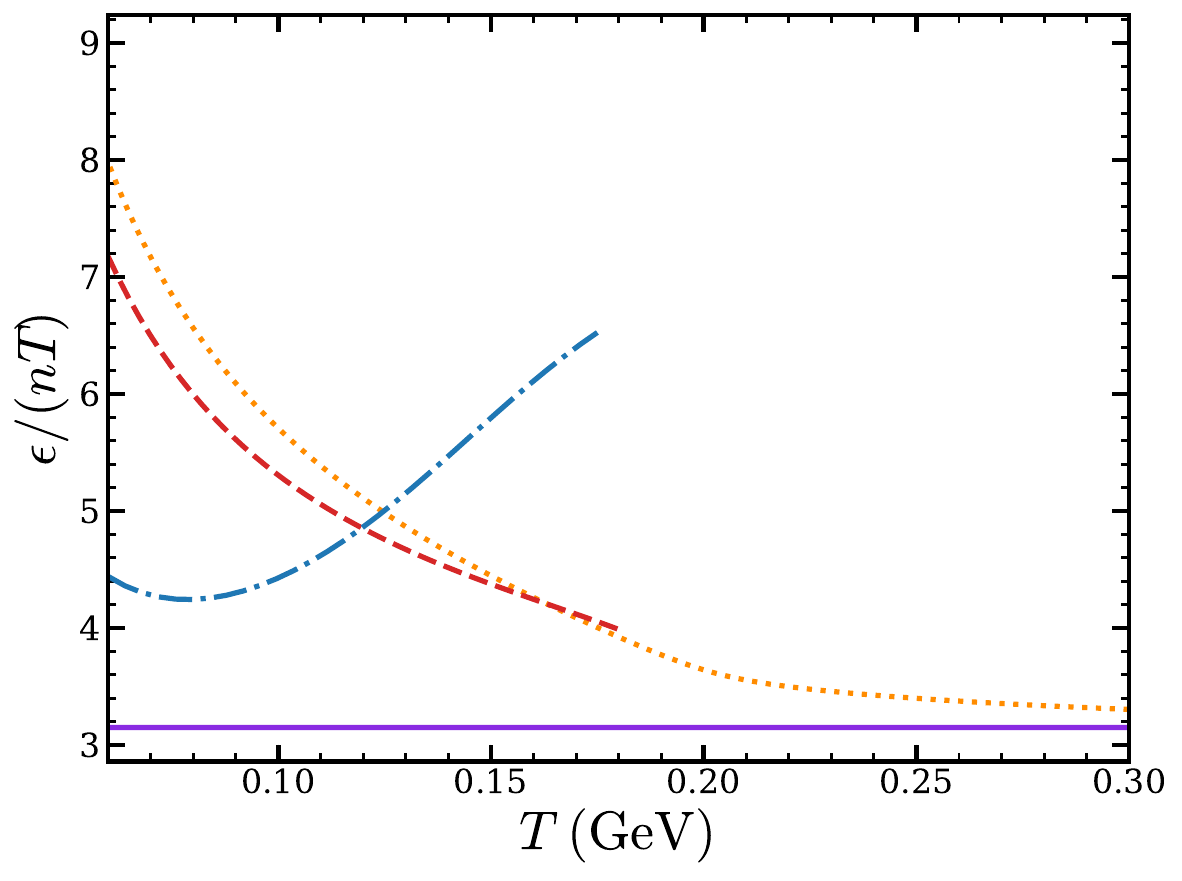}
	\end{subfigure}
	
	\caption{(Color online) Top: normalized pressure $P/T^4$ and energy density $\varepsilon/T^4$ vs. $T$. 
		Middle: total number density $n/T^3$ and entropy density $s/T^3$ vs. $T$. 
		Bottom: dimensionless ratios $P/(nT)$ and $\varepsilon/(nT)$ vs. $T$. 
		Results from the HRG, NJL, and chiral effective models are compared with the massless three-flavor quark limit and the lattice QCD estimates \cite{Borsanyi:2021sxv}.}
	\label{thermo_T}
\end{figure}

\begin{figure}[htbp]
	\centering
	
	\begin{subfigure}{0.24\textwidth}
		\centering
		\includegraphics[width=\linewidth]{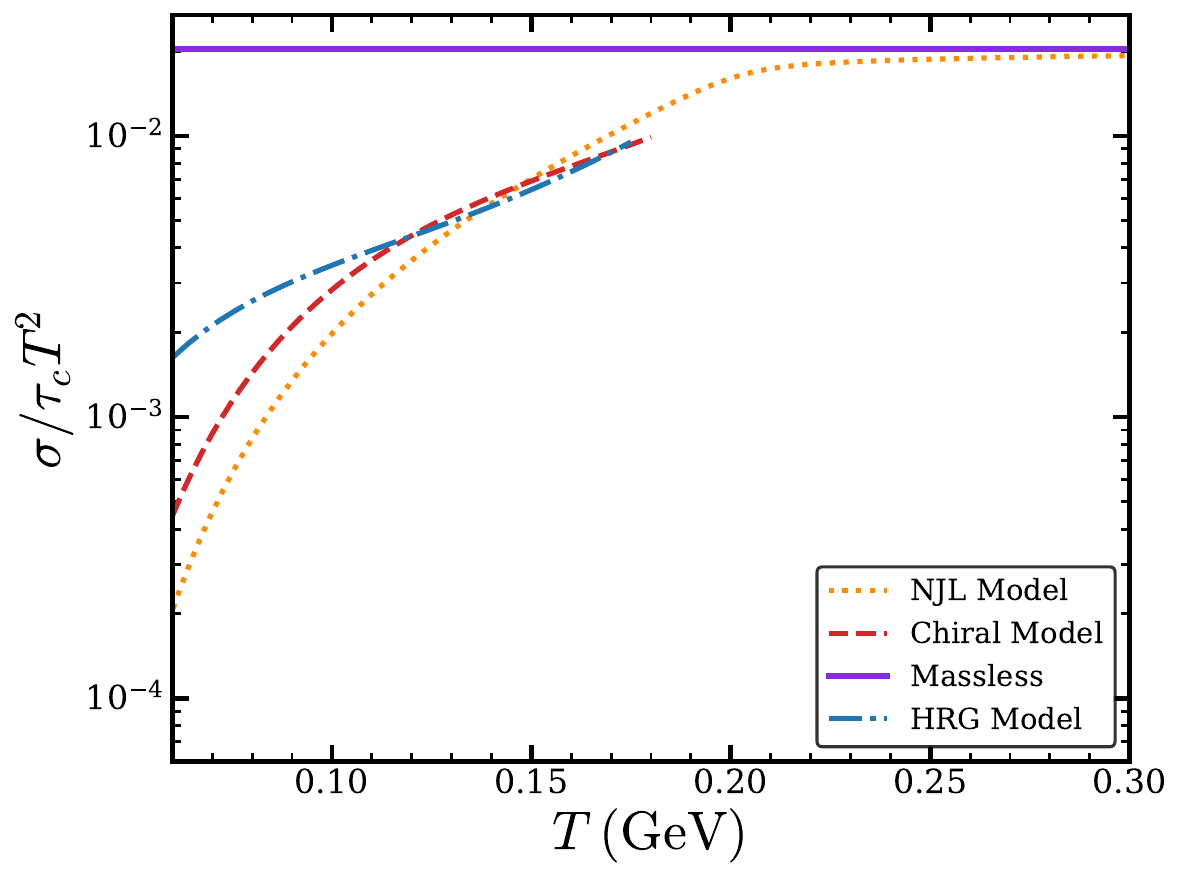}
	\end{subfigure}
	\hfill
	\begin{subfigure}{0.24\textwidth}
		\centering
		\includegraphics[width=\linewidth]{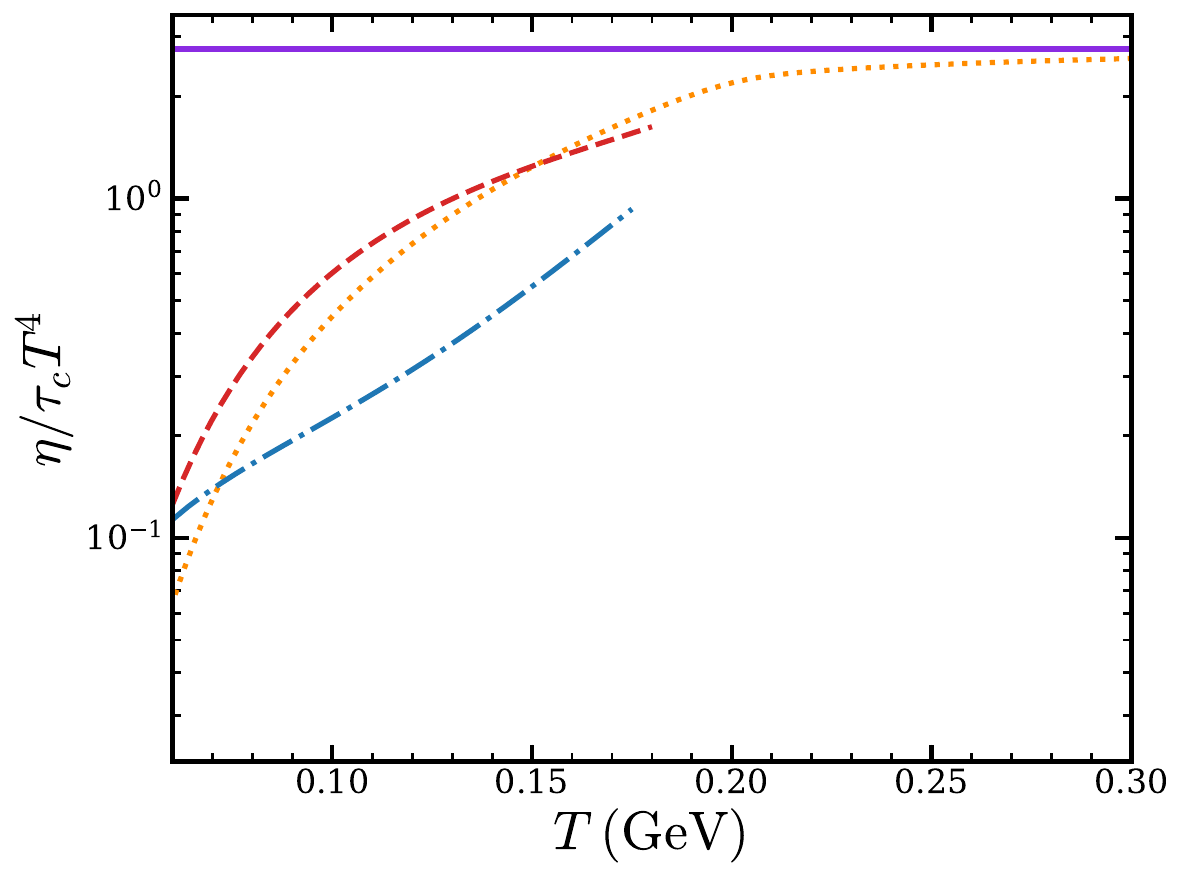}
	\end{subfigure}
	\caption{(Color online) Normalized electrical conductivity $\sigma/(\tau_c T^2)$ and shear viscosity $\eta/(\tau_c T^4)$ vs. $T$ in the HRG, NJL, and chiral effective models, compared with the corresponding massless three-flavor quark results.}
	\label{transport_T}
\end{figure}

\begin{figure}[htbp]
	\centering
	
	\begin{subfigure}{0.24\textwidth}
		\centering
		\includegraphics[width=\linewidth]{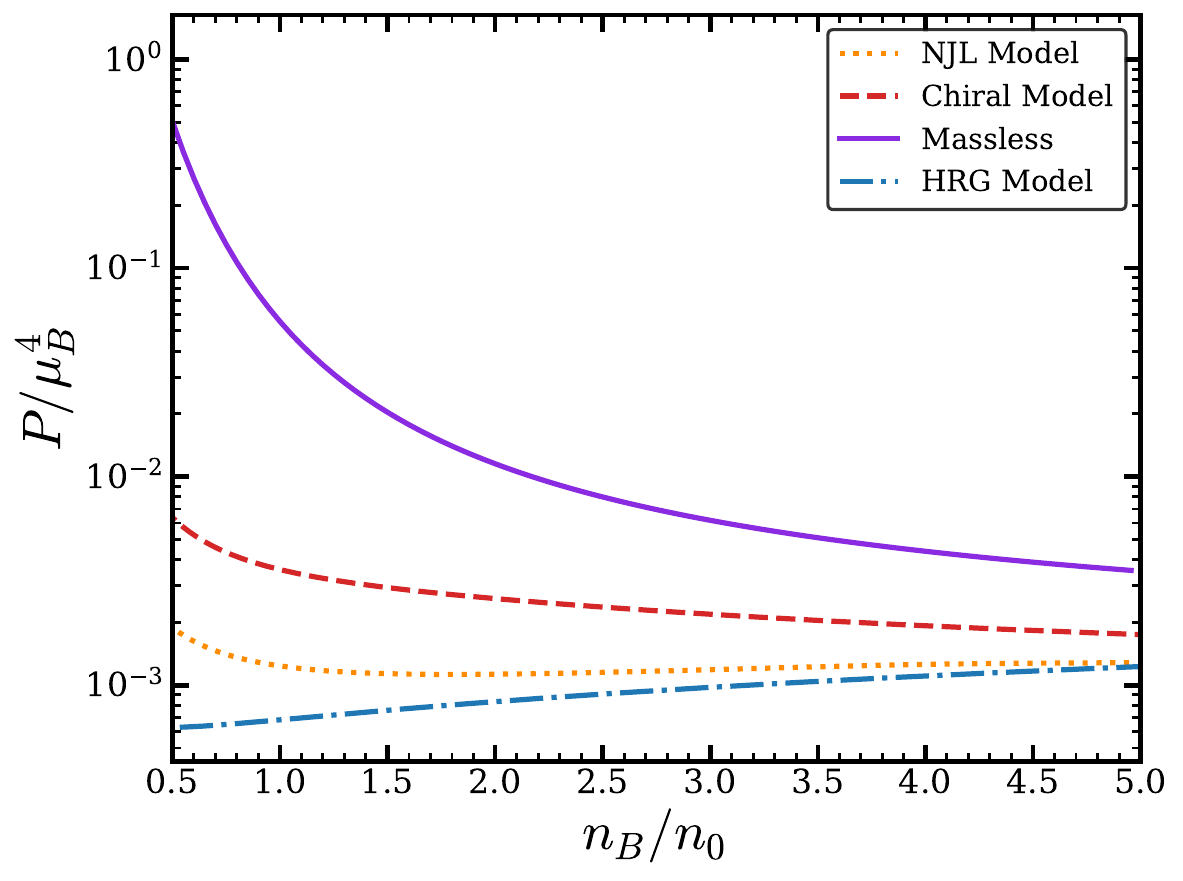}
	\end{subfigure}
	\hfill
	\begin{subfigure}{0.24\textwidth}
		\centering
		\includegraphics[width=\linewidth]{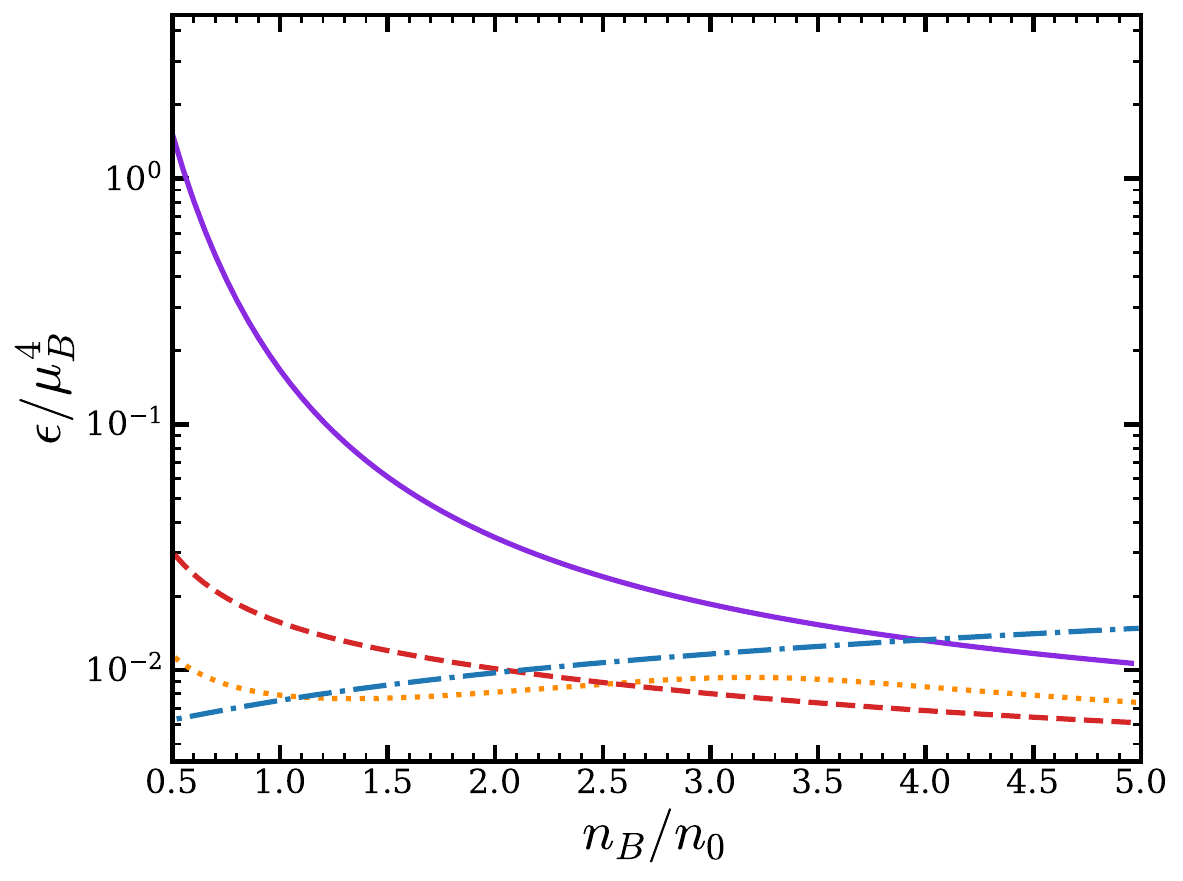}
	\end{subfigure}
	
	\vspace{0.5cm}
	
	\begin{subfigure}{0.24\textwidth}
		\centering
		\includegraphics[width=\linewidth]{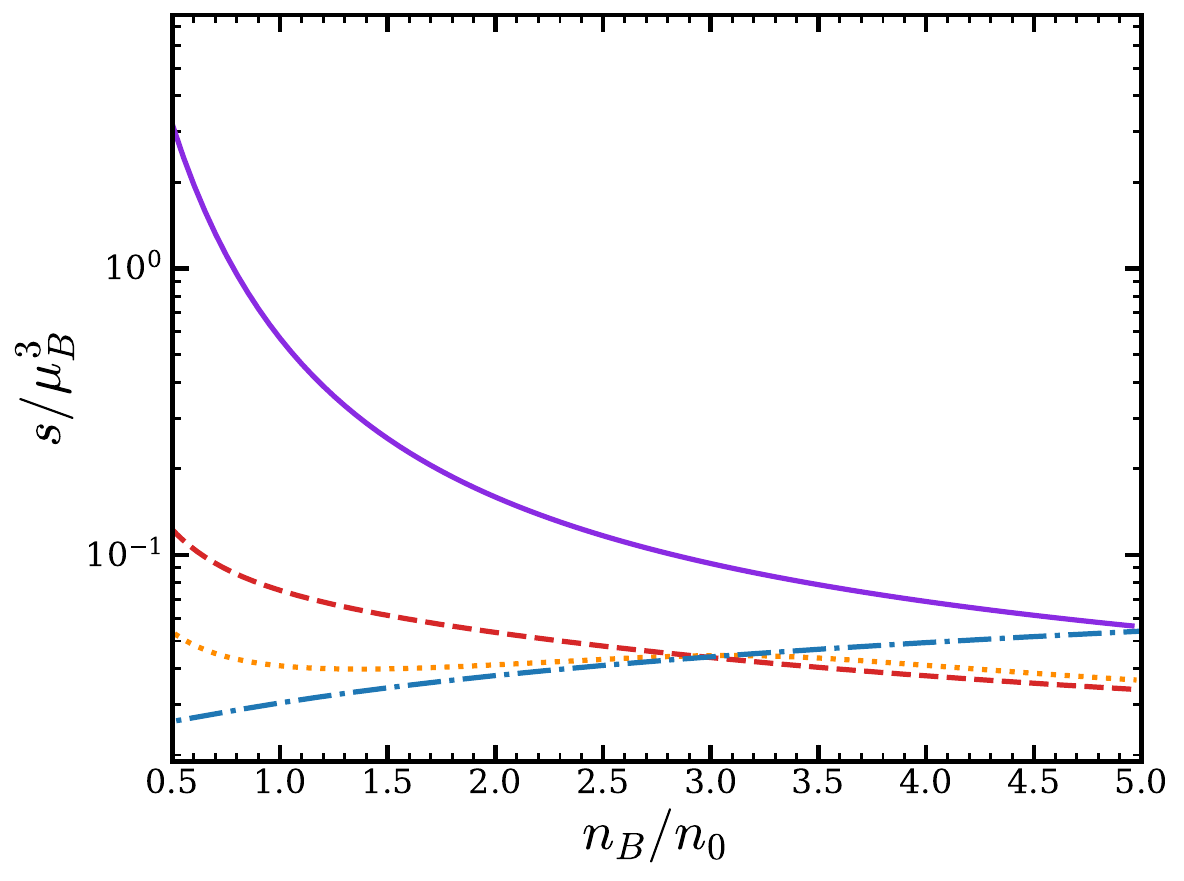}
	\end{subfigure}
	\hfill
	\begin{subfigure}{0.24\textwidth}
		\centering
		\includegraphics[width=\linewidth]{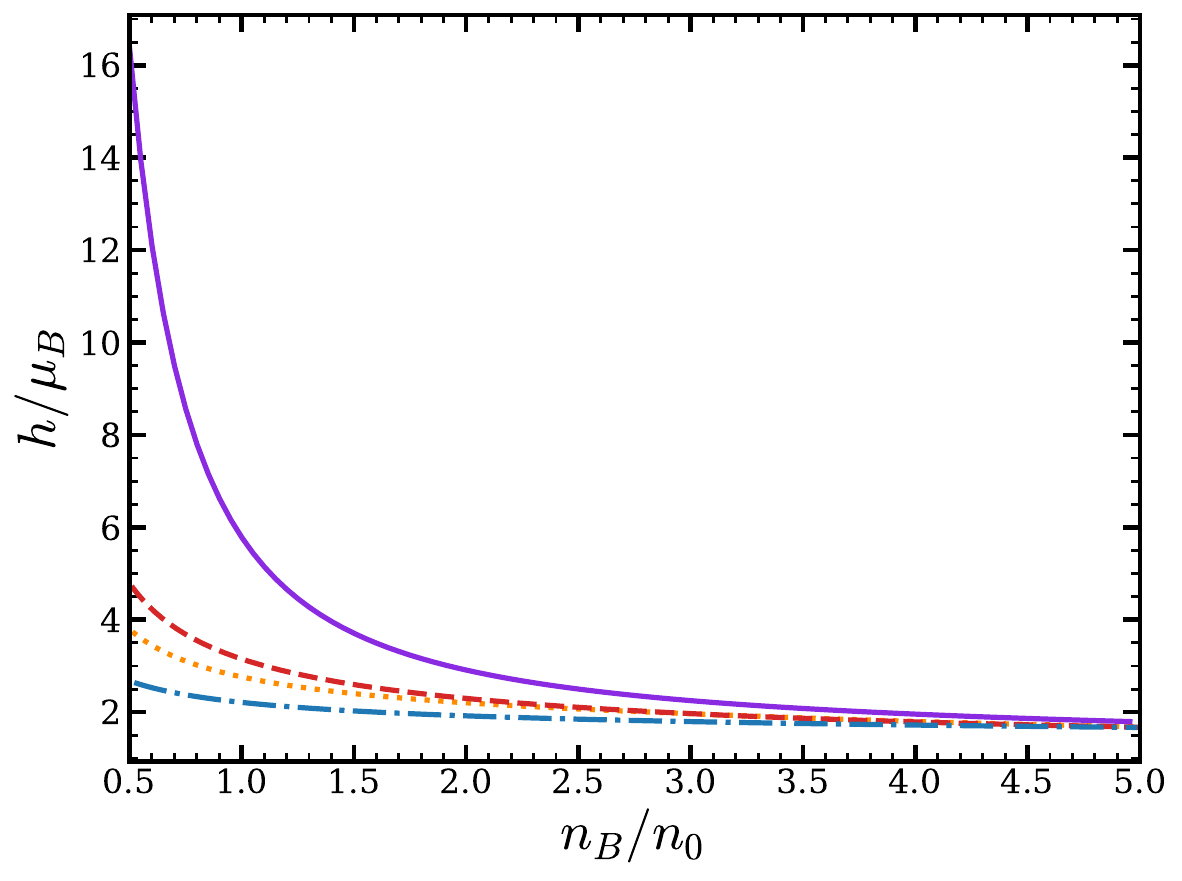}
	\end{subfigure}
	
	\vspace{0.5cm}
	
	\begin{subfigure}{0.24\textwidth}
		\centering
		\includegraphics[width=\linewidth]{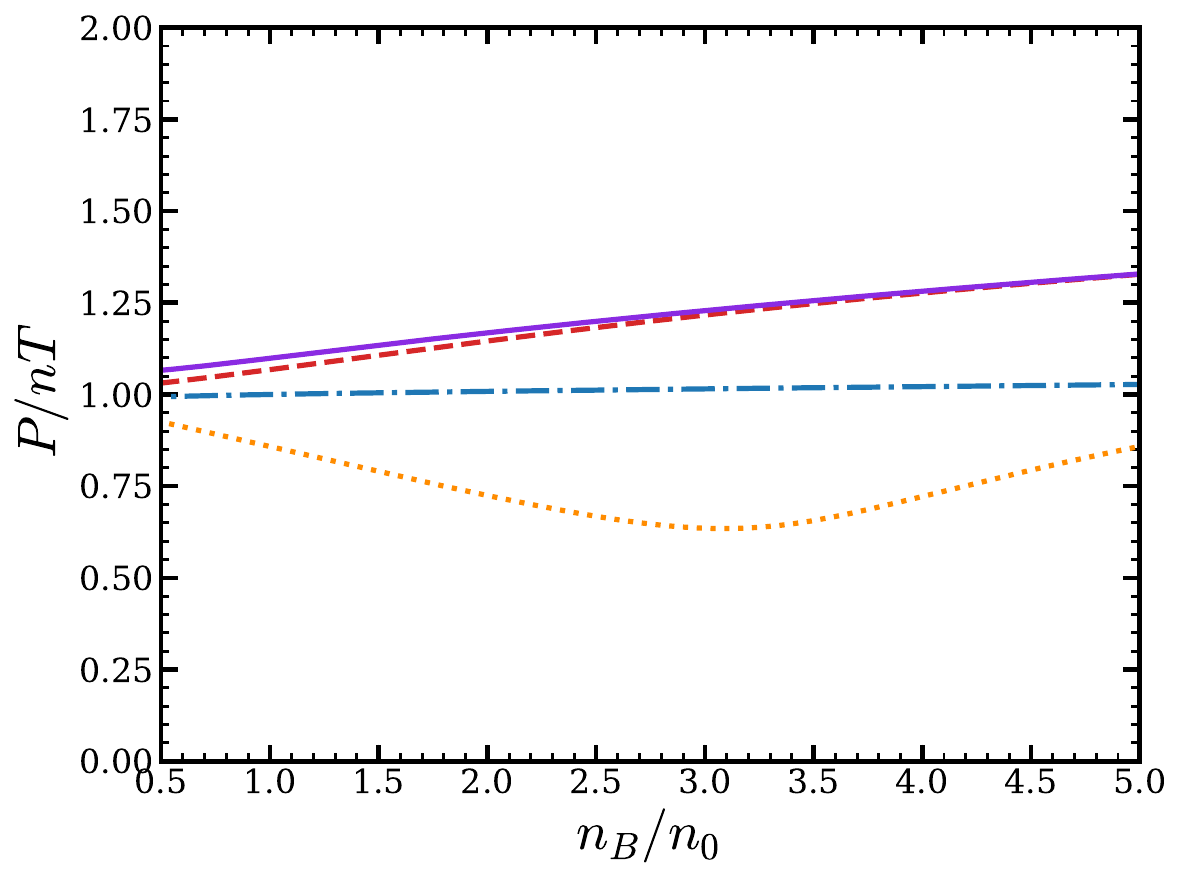}
	\end{subfigure}
	\hfill
	\begin{subfigure}{0.24\textwidth}
		\centering
		\includegraphics[width=\linewidth]{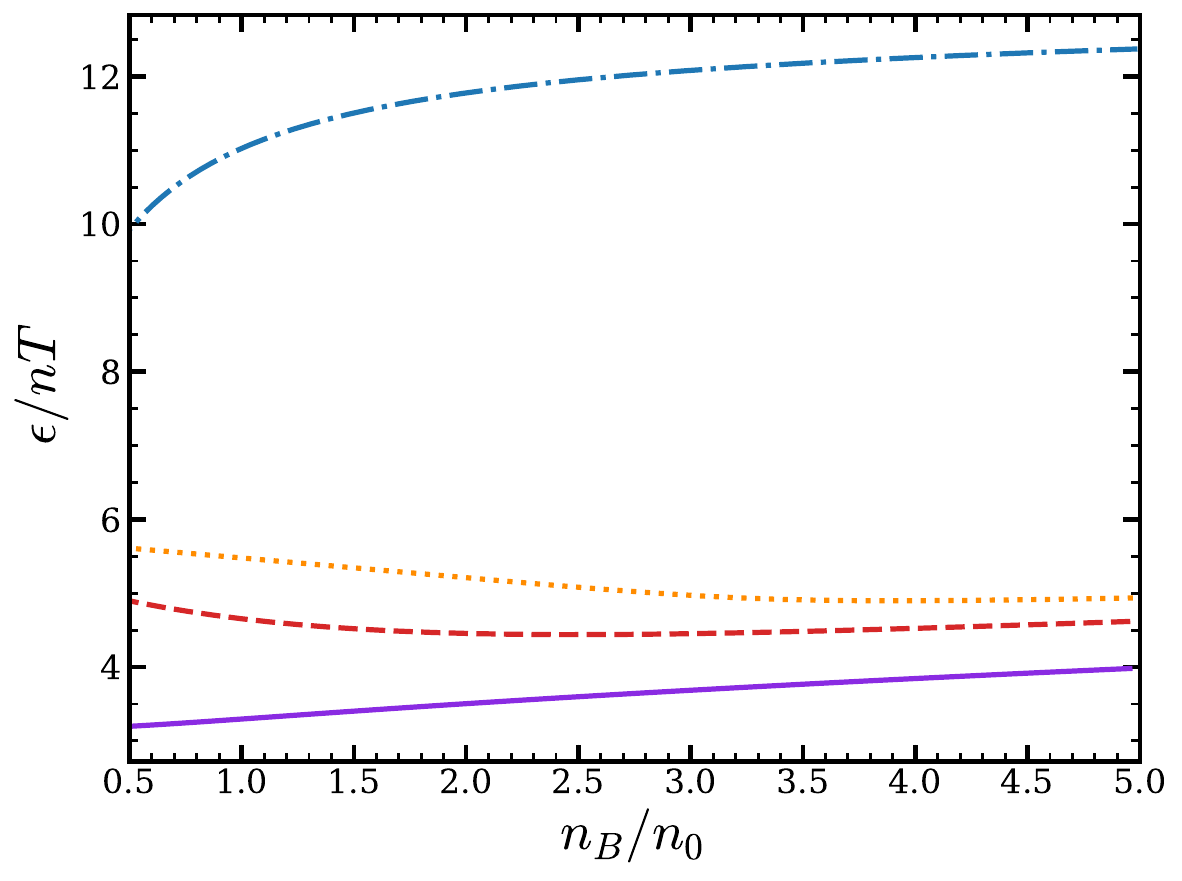}
	\end{subfigure}
	
	\caption{(Color online) Top: normalized pressure $P/\mu_B^4$ and energy density $\varepsilon/\mu_B^4$ vs. $n_{B}/n_{0}$. Middle: entropy density $s/\mu_B^3$ and enthalpy per baryon number $h/\mu_B$ vs. $n_B/n_0$. Bottom: dimensionless ratios $P/nT$ and $\varepsilon/nT$ vs. $n_B/n_0$. Results from the HRG, NJL, and chiral effective models are compared with the massless three-flavor quark limit at T=0.1 GeV.}
	\label{thermo_n}
\end{figure}

\begin{figure}[htbp]
	\centering
	
	\begin{subfigure}{0.24\textwidth}
		\centering
		\includegraphics[width=\linewidth]{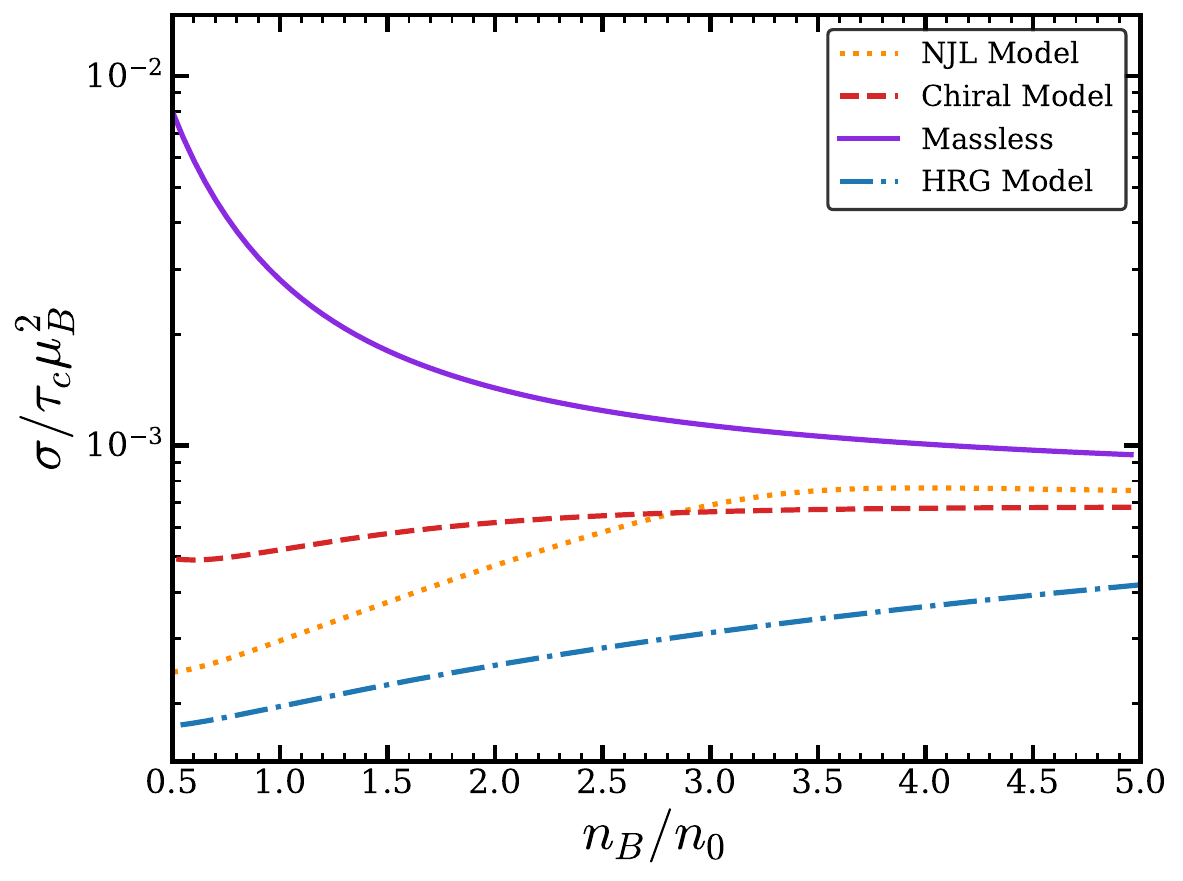}
	\end{subfigure}
	\begin{subfigure}{0.24\textwidth}
		\centering
		\includegraphics[width=\linewidth]{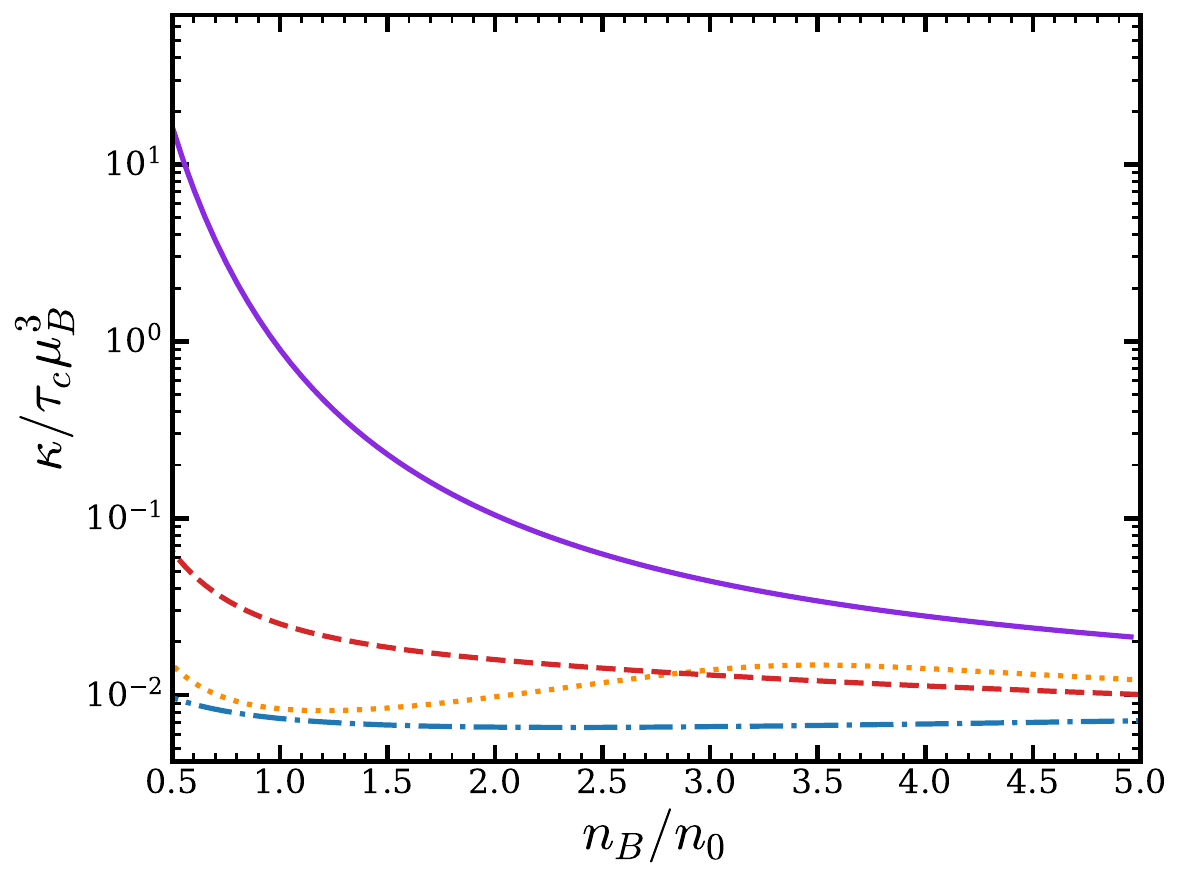}
	\end{subfigure}
	
	\vspace{0.5cm}
	
	\begin{subfigure}{0.24\textwidth}
		\centering
		\includegraphics[width=\linewidth]{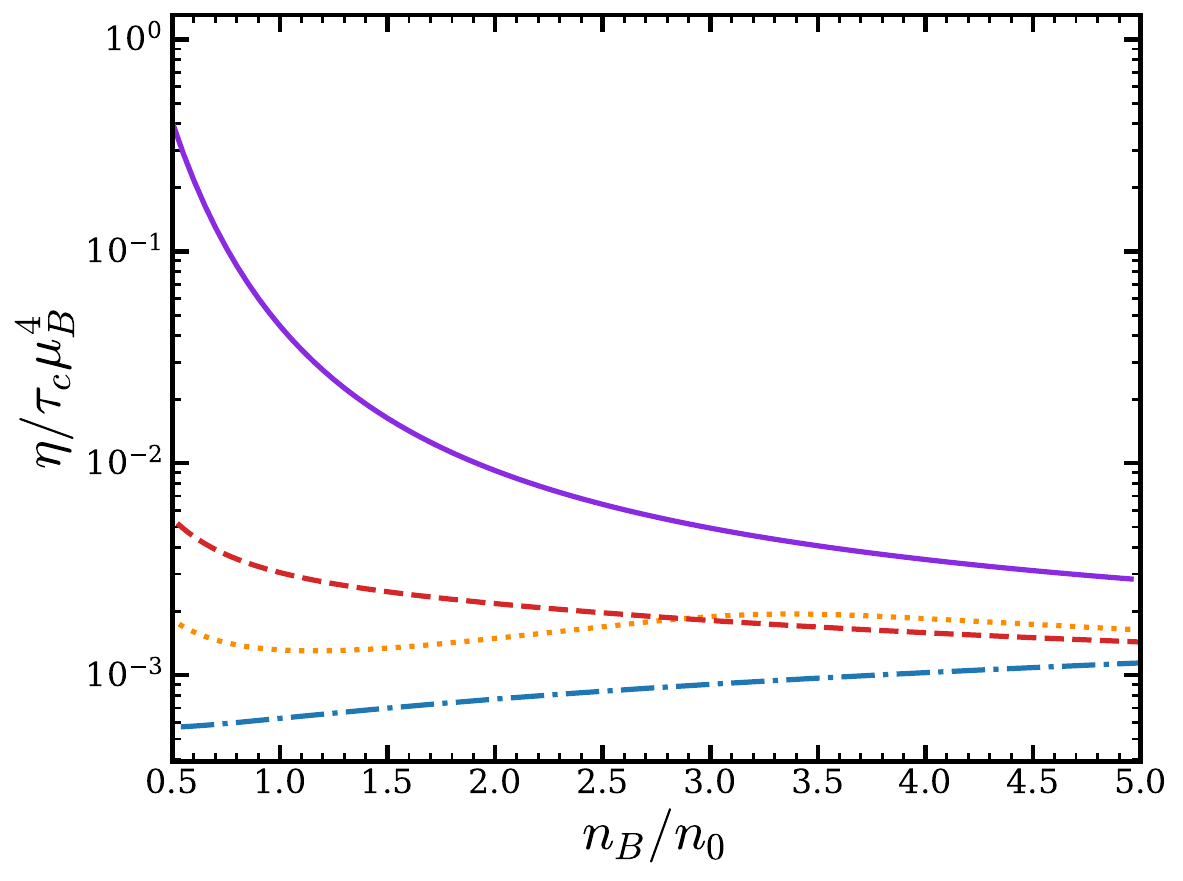}
	\end{subfigure}
	\begin{subfigure}{0.24\textwidth}
		\centering
		\includegraphics[width=\linewidth]{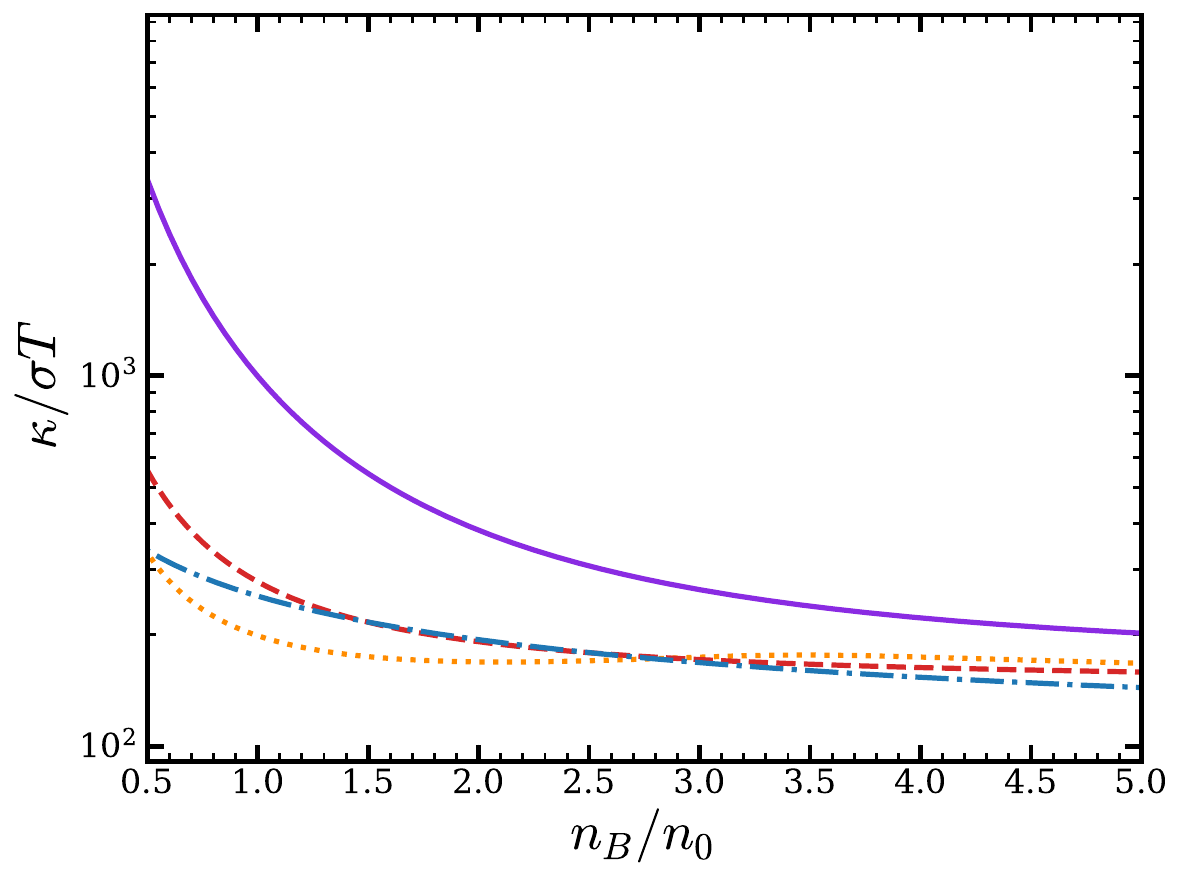}
	\end{subfigure}
	
	\caption{(Color online) Normalized electrical conductivity $\sigma/(\tau_c\mu_B^2)$, thermal conductivity $\kappa/(\tau_c\mu_B^3)$ and shear viscosity $\eta/(\tau_c\mu_B^4)$ as a function of scaled baryon density $n_B/n_0$ in HRG, NJL and chiral effective models compared with the corresponding values for a massless three-flavor quark system at T=0.1 GeV. The dimensionless ratio $\kappa/(\sigma T)$ is also plotted for all the cases.}
	\label{transport_n_T}
\end{figure}

\begin{figure}
	\centering
	\includegraphics[width=0.8\linewidth]{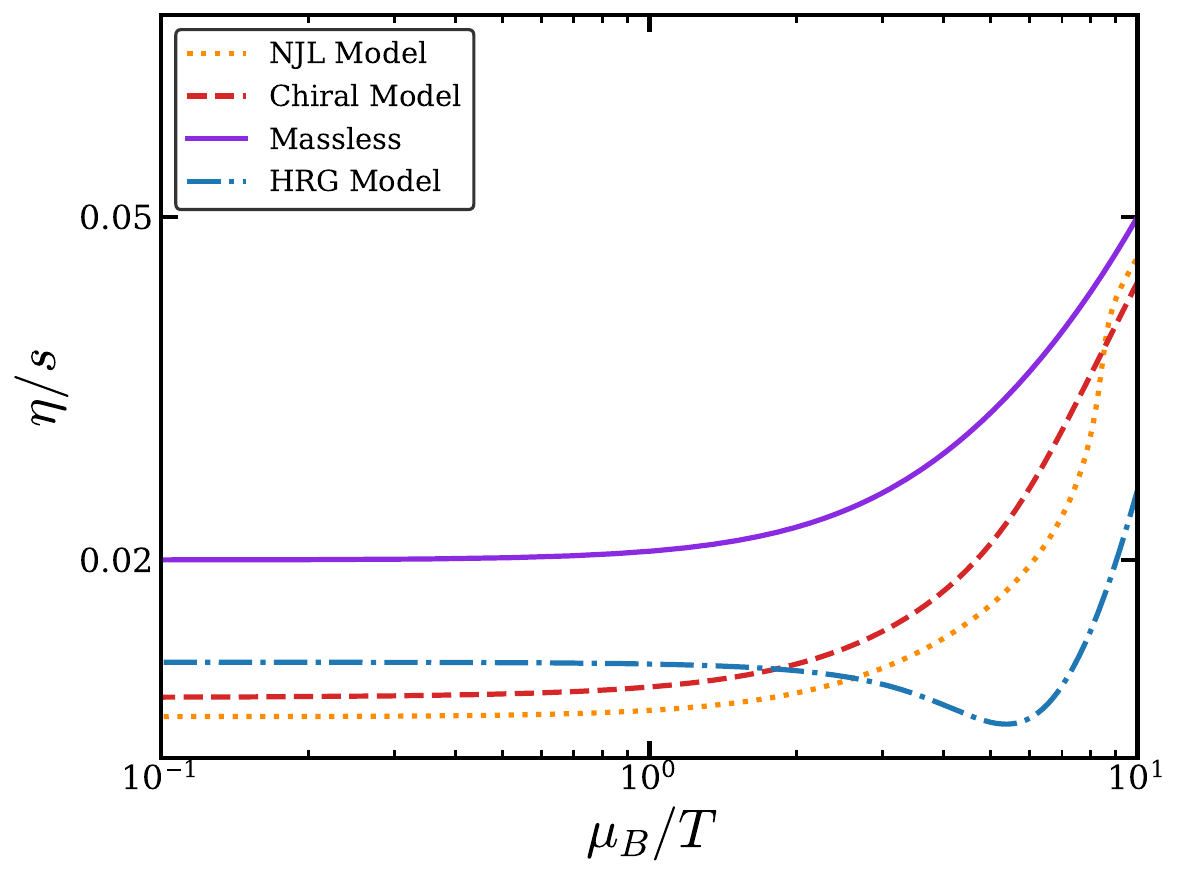}
	\caption{(Color online) Shear viscosity to entropy density ratio $\eta/s$ as a function of $\mu_B/T$ in HRG, NJL, and chiral effective models compared with the corresponding values for a massless three-flavor quark system at T=0.1 GeV. A constant relaxation time of 1 fm is assumed throughout the entire density range.}
	\label{eta}
\end{figure} 

Fig.~\ref{thermo_T} shows the temperature dependence of thermodynamic quantities at zero baryon density or chemical potential in the HRG, NJL, and chiral effective models along with the calculations for a massless three-flavor quark system often referred to as the Stefan-Boltzmann (SB) limit. For a massless system, in the limit $\mu_{B}\rightarrow 0$, the thermodynamic quantities can be expressed in terms of the Riemann zeta function $\zeta(s)$. For a three-flavor quark system, this turns out to be,
\begin{align}
	n &= \frac{27}{\pi^2}\zeta(3)T^3 =\frac{27}{\pi^2}1.202~ T^3~,\\
	\varepsilon &= 3\frac{63}{2\pi^2}\zeta(4)T^4 =\frac{21}{20}\pi^2 T^4~,\\
	P &= \frac{63}{2\pi^2}\zeta(4)T^4 =\frac{7}{20}\pi^2T^4~,\\
	s &= \frac{126}{\pi^2}\zeta(4)T^3 =\frac{7}{5}\pi^2T^3~,	
\end{align}
since $\int_0^\infty \frac{x^s}{e^x+1}dx=(1-2^{-s})\Gamma(s+1)\zeta(s+1)$. So, for massless quark matter, we see that the quantities energy density $\varepsilon$ and pressure $P$ vary as $T^4$, whereas the entropy density $s$ and total baryon density $n$ vary as $T^3$ as a function of temperature. Normalized energy density $\varepsilon/T^4$, pressure $P/T^4$, total density $n/T^3$, and entropy density $s/T^3$ are plotted, where the results for massless quarks provide the upper limit for these quantities and lie on a horizontal line parallel to the x-axis (purple solid). The results for other models, HRG (blue dash-dotted), NJL (yellow dotted), and the chiral effective model (red dashed),  are plotted in Fig.~\ref{thermo_T} along with the lattice QCD estimates \cite{Borsanyi:2021sxv}. Below the transition temperature, within the hadronic temperature range, the values of all thermodynamical quantities estimated from HRG, NJL, and chiral effective model are quite lower than their massless value (chiral or SB limit). This suppression is a well-known non-perturbative QCD effect supported by the LQCD simulations~\cite{Borsanyi:2013bia,Karsch:2001vs,HotQCD:2014kol}. The NJL results approach the massless limit at high temperature, consistent with the restoration of chiral symmetry above the transition temperature. Since the HRG model has hadronic degrees of freedom, the results are plotted only in the hadronic region. The chiral model is also more appropriate in low temperature and high density domains, so for the zero net baryon density case, we have kept their results within the low temperature domain.

In an attempt to quantify the thermodynamic properties, we also plot the quantities $\varepsilon/nT$ and $P/nT$, which are reminiscent of the equi-partition law and the ideal gas equation of state. The massless limit for these quantities is 3.15 and 1.05, respectively. These values are exactly 3 and 1 for a classical ideal gas, which follows the Maxwell-Boltzmann distribution. The equi-partition relation then interprets the average energy per particle as $\langle\varepsilon\rangle=\varepsilon/n=3T$, where 3 can be considered as the degrees of freedom. Here, for a massless QGP, we obtain this value to be around 3.15, and interestingly, this value of effective degrees of freedom changes with temperature in the effective QCD models. The results for effective quark models, NJL, and chiral effective models approach this value for $\varepsilon/nT$ and $P/nT$ as we approach the chiral limit at high temperatures due to chiral symmetry restoration. However, the HRG model results exhibit an exceptional trend. This deviation from the expected chiral-limit behavior likely arises from the underlying philosophy of the HRG framework, which treats hadrons as non-interacting degrees of freedom and does not have explicit chiral symmetry restoration.

Similarly, the transport coefficients at zero net baryon density are well-studied, and here we present the results for electrical conductivity and shear viscosity at finite temperature for reference before moving on to the results at finite density. Fig.~\ref{transport_T} displays the normalized electrical conductivity $\sigma/(\tau_c T^2)$ and shear viscosity $\eta/(\tau_c T^4)$ of a three-flavor quark system in the massless limit (purple solid line) as well as in the NJL (yellow dotted curve) and chiral effective (red dashed curve) models. Results for the HRG model (blue dash-dotted curve) are also included.

Similar to the thermodynamic quantities, the massless estimates can be a good reference point for different model estimations of transport coefficients, and the results of NJL and chiral effective model should approach these values as the chiral symmetry is restored. The massless limit for the transport coefficients, electrical conductivity, and shear viscosity has the following forms:
\begin{align}
	\sigma &= \frac{4e^2}{3\pi^2}\tau_c~ \zeta(2)T^2 =\frac{2e^2}{9}\tau_c T^2,\\
	\eta &= \frac{126}{5\pi^2}\tau_c~ \zeta(4)T^4 =\frac{7\pi^2}{25}\tau_c T^4,
\end{align}
which suggests that the normalized electrical conductivity and shear viscosity are constants, and they can serve as reference lines like SB or chiral limit for the transport coefficients. The results for HRG, NJL, and chiral effective models are lower than this value at lower temperatures but increase as the temperature approaches the massless limit. 

After providing a benchmark for thermodynamic quantities and transport coefficients at finite temperature and zero baryon density, we now outline the deviation we obtain in these quantities at finite net baryon density, which is the core aim of the present article. This can be particularly relevant to the upcoming CBM and NICA colliders, where nuclear matter at finite baryon density is expected.

Fig.~\ref{thermo_n} shows the density dependence of the thermodynamic quantities. Normalized pressure $P/\mu_B^4$, energy density $\varepsilon/\mu_B^4$, (top panel) entropy density $s/\mu_B^3$, and enthalpy per baryon number $h/\mu_B$ (middle panel) are plotted as a function of scaled baryon density $n_B/n_0$, where $n_0=0.15~$fm$^{-3}$ is the nuclear saturation density.  Again, the results are presented for HRG, NJL, and chiral effective models along with the corresponding values for a massless three-flavor quark system at a constant temperature of 0.1 GeV. For the massless case, increasing the density drives the system into the degenerate gas regime ($\mu_{B}/T \gg 1$), where the $T \to 0$ limit applies and the Fermi--Dirac distribution reduces
to a step function. Analogous to the $\mu_{B} \to 0$ limit, where
$P,\ \epsilon \propto T^{4}$ and $n,\ s \propto T^{3}$, the degenerate ($T \to 0$) limit yields
$P,\ \epsilon \propto \mu_{B}^{4}$, $s \propto \mu_{B}^{3}$, and $h \propto \mu_{B}$. We have plotted normalized thermodynamical quantities ($P/\mu_{B}^4,\epsilon/\mu_{B}^4,s/\mu_{B}^3,h/{\mu_{B}}$) to demonstrate their approach to the degenerate gas limit at high baryon density ($n_B>4n_0$). The values for the massless quark system provide the upper bound for these values, and the corresponding values in NJL and chiral effective models approach this value as density increases. The merging of NJL and chiral model estimation with massless case results indicates the possibility of chiral symmetry restoration at high density ($n_B>4n_0$). So, if CBM and NICA matter reach such a high density at $T=0.1$ GeV, then both the chiral limit and the degenerate gas limit can be achieved there. HRG model calculation does not follow this trend, particularly at high densities. This can be attributed to the hadronic degrees of freedom considered in the HRG model, which lacks the chiral symmetry philosophy. The dimensionless ratios $P/(nT)$ and $\varepsilon/(nT)$ for all cases are shown in the bottom panel of Fig.~\ref{thermo_n}. Additional $P/{(n\mu_{B})}$ and $\epsilon/{(n\mu_{B})}$ (alternate ratio) are also plotted in appendix Fig.~\ref{appendix2}. These ratios at finite baryon chemical potential should be compared with the $\mu_{B}=0$ results presented in the bottom panel of Fig.~\ref{thermo_T}. In contrast to the straight lines obtained at $\mu_{B}=0$, the massless results now increase with the baryon density. At $\mu_{B} \rightarrow 0$, small deviations of EOS from $P=nT$ and $\epsilon=3nT$ are due to Riemann-Zeta (quantum) factor. When $\mu_{B}$ or $n_B$ increases, these quantum factors increase through the Fermi integral function. Nevertheless, to highlight the similarity with Fig.~\ref{thermo_T}, we note that the chiral and NJL results converge toward the massless quark results at high densities. For the ideal–gas quotient $P/(nT)$ (equal to one for an ideal gas), the massless results act as an upper bound, whereas for the equipartition quotient $\varepsilon/(nT)$ (equal to three for an ideal gas), the massless curve serves as a lower bound—consistent with the behavior observed at $\mu_{B}=0$. Let us wrap up the discussion on thermodynamic variables of the dense quark matter by stating that similar behavior can be observed by plotting the quantities against $\mu_{B}/T$ instead of $n_{B}/n_{0}$. This becomes apparent as one realizes the baryon density as well as the total number density both increase monotonically with an increase in $\mu_{B}/T$ ( see Fig.~\ref{appendix1} of the appendix). The dimensionless ratio $\frac{\mu}{T}$, sometimes referred to as the reduced chemical potential in the context of graphene and other condensed-matter many-body systems, provides an interesting physical perspective. In the graphene system, the Fermi energy of electrons plays the role of chemical potential in the FD distribution, and it can be tuned by doping the system. Two domains are identified in graphene systems: the Dirac field (DF) domain
($\mu/T \ll 1$) and the Fermi liquid (FL) or Fermi gas (FG) domain
($\mu/T \gg 1$). The former exhibits Poiseuille electron flow, a hallmark of
electron hydrodynamics~\cite{Kumar_2022,Gugnani_2025,srivastava2025electronhydrodynamicsviscositytensor},
whereas the latter is characterized by ohmic transport. Motivated by this
analogy, a similar domain classification for massless quark matter, quark
matter with constituent masses (within NJL or chiral models), and HRG matter, requires detailed knowledge of the total and net baryon densities as functions of $\mu_{B}/T$ (see Fig.~\ref{appendix1} of the appendix).

We now shift our attention to the behavior of the transport coefficients at finite baryon density. In Fig.~\ref{transport_n_T} and \ref{eta}, the variation of transport coefficients with net baryon density has been presented. Transport coefficients, unlike the thermodynamic variables, are determined by two factors--a relaxation time part and a thermodynamic phase-space part \cite{Dwibedi:2024mff}. To emphasize the thermodynamical phase-space part of all the transport coefficients, we have plotted the normalized electrical conductivity $\sigma/(\tau_c\mu_B^2)$, thermal conductivity $\kappa/(\tau_c\mu_B^3)$, and shear viscosity $\eta/(\tau_c\mu_B^4)$ as a function of scaled baryon density $n_B/n_0$ in HRG, NJL and chiral effective models along with the corresponding values for a massless three-flavor quark system. These normalizations for the massless case at high $n_B$ provide saturation (horizontal) values of different transport coefficients, which may be considered as a reference line or an upper bound, as done by the SB limit for $\mu_{B}\rightarrow 0$ picture. The normalized value of electrical conductivity, thermal conductivity, and shear viscosity all differ significantly from the corresponding massless results for $n_{B}/n_{0}\leq 5$, suggesting a significant non-perturbative contribution in this domain. In our estimation, these non-perturbative effects are included in the density and temperature-dependent quark masses derived in the NJL and chiral model framework. On the other hand, the HRG model encodes these non-perturbative effects through effective hadronic degrees of freedom. 

Certain ratios involving thermodynamic and transport parameters play a vital role in characterizing the properties of a many-body system. In this context, the ratios $\kappa/(\sigma T)$ and $\eta/s$ are presented to examine their behavior in baryon-rich environments relevant for future collider facilities such as CBM and NICA, and to compare them with the corresponding values for net-baryonless QCD matter produced at LHC and top RHIC energies. The behavior of $\eta/s$ provides valuable insight into the underlying scattering mechanisms and the various phases of the system, particularly in connection with phase transitions. For example, systems undergoing a liquid to gas phase transition—such as water, nitrogen, or helium—exhibit a characteristic valley-shaped structure in $\eta/s$ with a dip near the critical temperature; a similar feature is observed in the QGP, as discussed in Ref. \cite{Csernai:2006zz}. In Fig.~\ref{eta}, we plot $\eta/s$ as a function of $\mu_B/T$ for different cases, assuming a constant relaxation time $\tau_c \approx 1$~fm in order to highlight the thermodynamic phase-space dependence of $\eta/s$. We observe that $\eta/s$ remains nearly constant in the region $\mu_B/T < 1$, while it increases toward larger values as $\mu_B/T$ enters the domain $\mu_B/T > 1$. This growing tendency in the domain $\mu/T > 1$ has also been observed in the context of graphene in Ref.~\cite{Aung:2023vrr}, where it was interpreted as a weakening of the strongly coupled nature of the system or a breakdown of its fluid properties. Similarly, for quark or hadronic degrees of freedom, the fluid behavior may deteriorate at high net baryon density.	Drawing an analogy from condensed matter physics-based many-body interaction effects, very small values of the relaxation time $\tau_c$ (i.e., a strongly coupled regime) in the domain $\mu/T \ll 1$ are responsible for the emergence of electron hydrodynamics (eHD). In contrast, the regime $\mu/T \gg 1$, corresponding to a degenerate electron gas where non-interacting fermions with strong Pauli blocking dominate, tends to lose fluid characteristics due to larger values of $\tau_c$ (i.e., a weakly coupled or nearly non-interacting regime). Inspired by the graphene system, one may therefore expect a breakdown of fluid behavior or the emergence of non-fluid many-body characteristics in (non-perturbative) quark or hadronic matter at high net baryon density, which can be systematically explored in future experimental facilities 
such as CBM and NICA. Notably, a close similarity has already been observed between graphene in the 
Dirac fluid regime ($\mu/T \ll 1$) and quark or hadronic matter near vanishing net baryon density, 
as produced in RHIC and LHC experiments. In both cases, the systems exhibit strong coupling and 
robust fluid behavior, despite belonging to many-body systems operating at vastly different energy 
scales (In graphene, the relevant temperature and Fermi energy lie in the eV range, whereas for 
quark or hadronic matter, they are of the order of MeV). RHIC and LHC matter ~\cite{Heinz:2013th,Gale:2013da} show the highest strongly coupled aspect as their $\frac{\eta}{s}$ values are quite close to its quantum lower bound or KSS bound $\frac{\eta}{s}=\frac{1}{4\pi}$~\cite{Kovtun:2004de}. In Fig.~\ref{eta}, this KSS line is shown by a horizontal black line, and we observe that by considering the same value of relaxation time ($\tau_{c}\approx 1$ fm), the NJL model estimation is closer to the KSS line than other model estimations. In principle, the relaxation time $\tau_c$ should also be calculated microscopically. However, a general consensus on the determination of relaxation time is still not reached particularly in the heavy ion collisions partly due to the lack of experimental signatures validating them. In the present work, since we use three different models to study the thermodynamic and transport properties we refrain from using any particular model of relaxation time and use $\tau_c$ = 1 fm as mentioned. An important component in the calculation of relaxation times in hot and dense matter is the Pauli suppression of the available phase space for collisions and there are several estimates for the same in dense matter including \cite{Hakim:1993zz,Mornas:1994cc,Danielewicz:1984kt} and references therein. However, our primary interest here is to focus on the thermodynamic phase-space contribution to $\eta/s$ within these models, in order to obtain a qualitative understanding of the transition from fluid to non-fluid behavior.  Here, we notice that the phase space of $\frac{\eta}{s}$ for all models shows a diverging trend at high density, which indicates a transition from strong to weak correlation, or fluid to non-fluid (weak correlation may fail to develop fluid velocity gradient). In a more realistic picture, the order of magnitude of $\tau_{c}$ may also diverge for higher densities if we take guidance from condensed matter (graphene) knowledge. Therefore, we anticipate a tendency toward a fluid-to-nonfluid transition as one moves from the RHIC/LHC environment, characterized by $\mu_{B}/T \ll 1$, to the 
CBM/NICA regime, where $\mu_{B}/T \gg 1$.
Likewise, the violation of the Wiedemann–Franz law~\cite{doi:10.1126/science.aad0343,Majumdar_2025} at the Dirac point in graphene has been argued to share the same underlying physics as that observed in the matter produced at the LHC, as noted in Ref.~\cite{Dwibedi:2024nni}. In what follows, we discuss the behavior of 
$\kappa/(\sigma T)$ in comparison with the electron–hole plasma in graphene, a system that exhibits striking similarities with the QGP. The normalized ratio of electric conductivity to thermal conductivity is of paramount importance for an electrically charged many-body system. It measures the ease of charge flow in comparison to heat flow in the system. In normal metals at room temperature, the ratio $\kappa_e/(\sigma_e T)$ is observed to be 
a constant, $L_0 = \frac{\pi^2}{3}\frac{k_B^2}{e^2}$, known as the Lorenz number~\cite{ashcroft1976solid}, where $k_{B}$ is the Boltzmann constant. 
The universality of the relation $\kappa_{e}/(\sigma_{e} T) = L_0$ for metals is well known as the 
Wiedemann--Franz law. It is worth pointing out that the metals are a degenerate electronic system at room temperature $k_{B}T\sim 25$ meV, whereas the Fermi energy $\epsilon_{F}=2-10$ eV. The Wiedemann-Franz law can be proved for metals assuming solid-like transport in this degenerate regime \cite{Win:2024gds}. Nevertheless, it turns out that this law can be violated in metals if one moves to the domain $\mu/T\leq 1$ by raising the temperature of the solid \cite{Win:2024gds}. Among condensed matter systems, the one most closely resembling the medium produced in heavy-ion collisions is likely graphene. Its charge carriers (electrons and holes) obey a relativistic energy–momentum dispersion relation and therefore form a quasi-relativistic plasma. Moreover, the electronic chemical potential $\mu$ in graphene can be tuned by adjusting the gate voltage, in close analogy to heavy-ion collision experiments where a QCD medium with finite $\mu_{B}$ can be created. A more detailed analogy between QGP and graphene can be found in \cite{Dwibedi:2024nni}. The variation of dimensionless ratios, such as $\kappa/\sigma T$, can shed light on the understanding of the properties of these two seemingly disparate physical systems: a two-dimensional quasi-relativistic electron-hole plasma formed in graphene and a three-dimensional relativistic QGP formed in heavy-ion collisions. In the bottom right panel of Fig.~\ref{transport_n_T}, we show the ratio $\kappa/\sigma T$ as a function of scaled baryon density. Assuming a solid-like transport, the value $\kappa/\sigma T$ for a single species massless system of unit charge $e$ is equal to $L_{0}=\frac{\pi^{2}}{3}\frac{k^{2}_{B}}{e^{2}}\approx 35.9$ in natural units, a number which is independent of the spatial dimension \cite{Dwibedi:2024nni}. In a multi-species system like the QGP, the above ratio should be modified to account for its degeneracy and the unequal magnitude of the electric charges of $u$,$d$ and $s$ quarks. For a 3-flavor quark matter with quark charges $e_{u}=2/3$ , $e_{d}=-1/3$ and  $e_{s}=-1/3$ the above ratio modifies to $\frac{\pi^{2}}{3}\frac{k^{2}_{B}}{e^{2}}\rightarrow \frac{N_{f}N_{c}}{N_{c}(e_{u}^{2}+e_{d}^{2}+e_{s}^{2})}\frac{\pi^{2}}{3}\frac{k^{2}_{B}}{e^{2}}\approx 161.4$. As baryon density increases, we observe in Fig.~\ref{transport_n_T} that $\kappa/\sigma T$ for the massless quark matter (purple line) approaches this value. Across all the values of baryon density, the massless curve serves as an upper bound for other model calculations. A severe violation of Wiedemann-Franz law, i.e., $\kappa/(\sigma T L_{0}) \gg 1$ is observed for $n_{B} \rightarrow 0$. This suggests that the Wiedemann-Franz law is violated for the baryon-less quark matter created in LHC and top RHIC energies. In the quark matter expected to be produced in the low-energy collisions at CBM and NICA, our observations suggest that the Wiedemann–Franz law is gradually restored as the collision energy decreases (or, equivalently, as $\mu_{B}$ and $n_{B}$ increase). A similar phenomenon has been observed experimentally \cite{doi:10.1126/science.aad0343,Majumdar_2025}—and explored theoretically \cite{Win:2024gds}—in graphene, where the Wiedemann–Franz law is recovered as one increases the gate voltage, or equivalently the electronic chemical potential. 
\section{Summary}
\label{Summary}	
Future heavy ion collision experiment facilities like CBM, FAIR in Germany, and NICA in Russia are the motivation for the present theoretical work. Here we have explored the thermodynamical and transport properties of quark and hadronic matter at finite (net) quark or baryon density, which is expected to be produced in those experimental facilities. In this study, we use three different effective models--the NJL model, the chiral effective model, and the HRG model. Initially, we briefly address the field theoretical models--the NJL model and the chiral effective model--to derive temperature and baryon chemical potential-dependent constituent quark masses from the quark condensates. The HRG model is then described, which encompasses all known hadrons and resonances to mimic the thermodynamics of hot and dense QCD matter effectively. Equipped with these three models, we write down the expressions for the thermodynamic quantities and transport coefficients. The thermodynamic quantities (and transport coefficients) are obtained by integrating over the equilibrium (and out-of-equilibrium) thermal distribution functions, incorporating medium-dependent masses in the NJL and chiral effective models, and by summing over hadronic degrees of freedom in the HRG model. The out-of-equilibrium thermal distribution function is obtained from the Boltzmann transport equation in relaxation time approximation. To begin with, we present the behavior of thermodynamic quantities and transport coefficients at zero baryon chemical potential as a function of temperature, and then discuss our main results at finite baryon density. 

At vanishing baryon chemical potential, a detailed understanding of various thermodynamical 
quantities such as pressure, energy density, and entropy density is available from LQCD 
calculations. A general conclusion from these studies is that their values remain lower than those 
of massless quark matter, i.e., below the Stefan-Boltzmann limit. The present work begins with 
qualitative consistency checks of this behavior using the NJL, chiral, and HRG models, and then 
extends the analysis to finite quark and baryon density scenarios. We also demonstrate deviations from the classical ideal (massless) gas equations of state, 
$P = nT$ and the equipartition relation $\epsilon = 3nT$, which arise due to quantum statistics 
and the nonperturbative interplay among these three models. Similar to the exploration of 
thermodynamical quantities in the $T$--$\mu$ plane, transport coefficients such as shear viscosity, 
electrical conductivity, and thermal conductivity are also analyzed. Their corresponding phase-space 
structures, obtained by normalizing with the relaxation time, exhibit comparable nonperturbative 
features in the $T$--$\mu$ plane.
Finally, we present two key ratios: (i) the shear viscosity to entropy density ratio 
$\eta/s$, and (ii) the thermal to electrical conductivity ratio $\kappa/(\sigma T)$. 
Both quantities exhibit a remarkable qualitative similarity with those observed in graphene. Although the characteristic energy scales of 
quark/hadronic matter and graphene differ significantly (MeV versus eV), both systems 
appear to exhibit a similar fluid-to-nonfluid transition pattern. In particular, both 
systems display fluid-like behavior in the regime $\mu/T \ll 1$, while nonfluid behavior 
emerges as $\mu/T \gg 1$. The reduced chemical potential $\mu/T$ can be tuned in quark/hadronic matter and in 
graphene through the center-of-mass energy of heavy-ion collisions and doping techniques, 
respectively. For graphene, earlier theoretical and experimental studies have established 
that:
\begin{itemize}
	\item[(i)] the ratio $\eta/s$ remains small and nearly saturated in the regime 
	$\mu/T \ll 1$, but exhibits a diverging trend as $\mu/T \gg 1$;
	\item[(ii)] the Wiedemann--Franz law is violated, i.e., $\kappa/(\sigma T) \neq L_0$, 
	in the regime $\mu/T \ll 1$, while it is restored, $\kappa/(\sigma T) = L_0$, in the 
	regime $\mu/T \gg 1$, signaling a transition from fluid-like to nonfluid behavior.
\end{itemize}
Our present analysis of quark and hadronic matter reveals qualitatively similar trends. 
However, a more detailed and systematic investigation will be required in the future to 
arrive at definitive conclusions regarding the existence and nature of a fluid-to-nonfluid 
transition in strongly interacting matter.

\section*{Data availability statement}
Data availability statement
This study has no associated datasets.
\section*{Code availability statement}
The code developed and used in
this study is publicly available at: \href{https://github.com/anand-rai-sahab/NJL-HRG-Chiral-Model}{Anand Rai GitHub Repository}
\section{Acknowledgement}
This work was supported in part by the Board of Research in Nuclear Sciences (BRNS) and the Department of Atomic Energy (DAE), Government of India, with Grant Nos. 57/14/01/2024-BRNS/313 (A.R. and S.G.), the Ministry of Education, Government of India  (D.R.M. and A.D.), the Ministry of Tribal Affairs, Government of India (P. M.), and Indian institute of technology Bhilai Innovation and Technology Foundation (IBITF) under the Tribal area Sub Plan (TSP), funded from Department of Science and Technology (DST), Government of India (R.S. and S.G.).

\section{Appendix}\label{appe1} 
\begin{figure}[htbp]
	\centering
	
	\begin{subfigure}{0.24\textwidth}
		\centering
		\includegraphics[width=\linewidth]{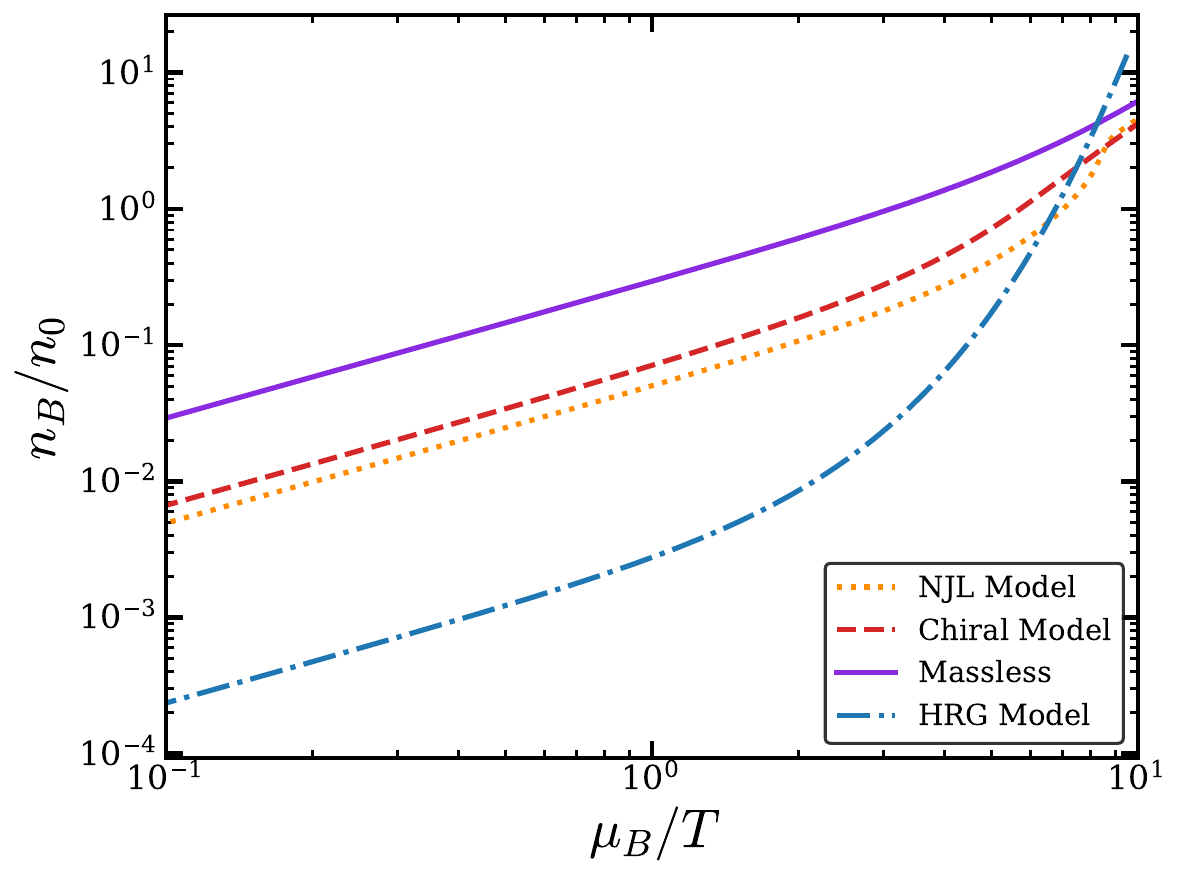}
	\end{subfigure}
	\hfill
	\begin{subfigure}{0.24\textwidth}
		\centering
		\includegraphics[width=\linewidth]{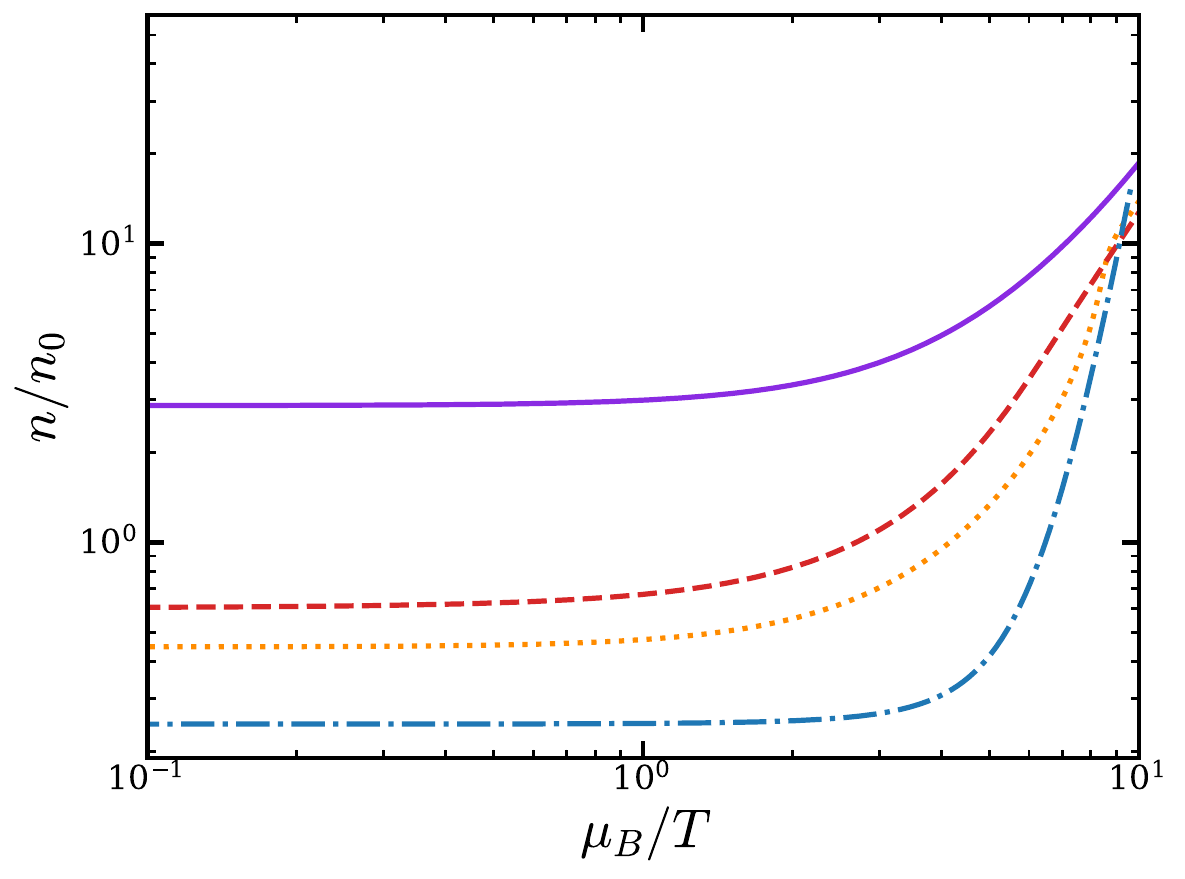}
	\end{subfigure}
	\caption{Total baryon density $n/n_0$ and net baryon density $n_B/n_0$ as a function of $\mu_B/T$ in HRG, NJL, and chiral effective models compared with the corresponding values for a massless three-flavor quark system.}
	\label{appendix1}
\end{figure}

\begin{figure}[htbp]
	\centering
	
	\begin{subfigure}{0.24\textwidth}
		\centering
		\includegraphics[width=\linewidth]{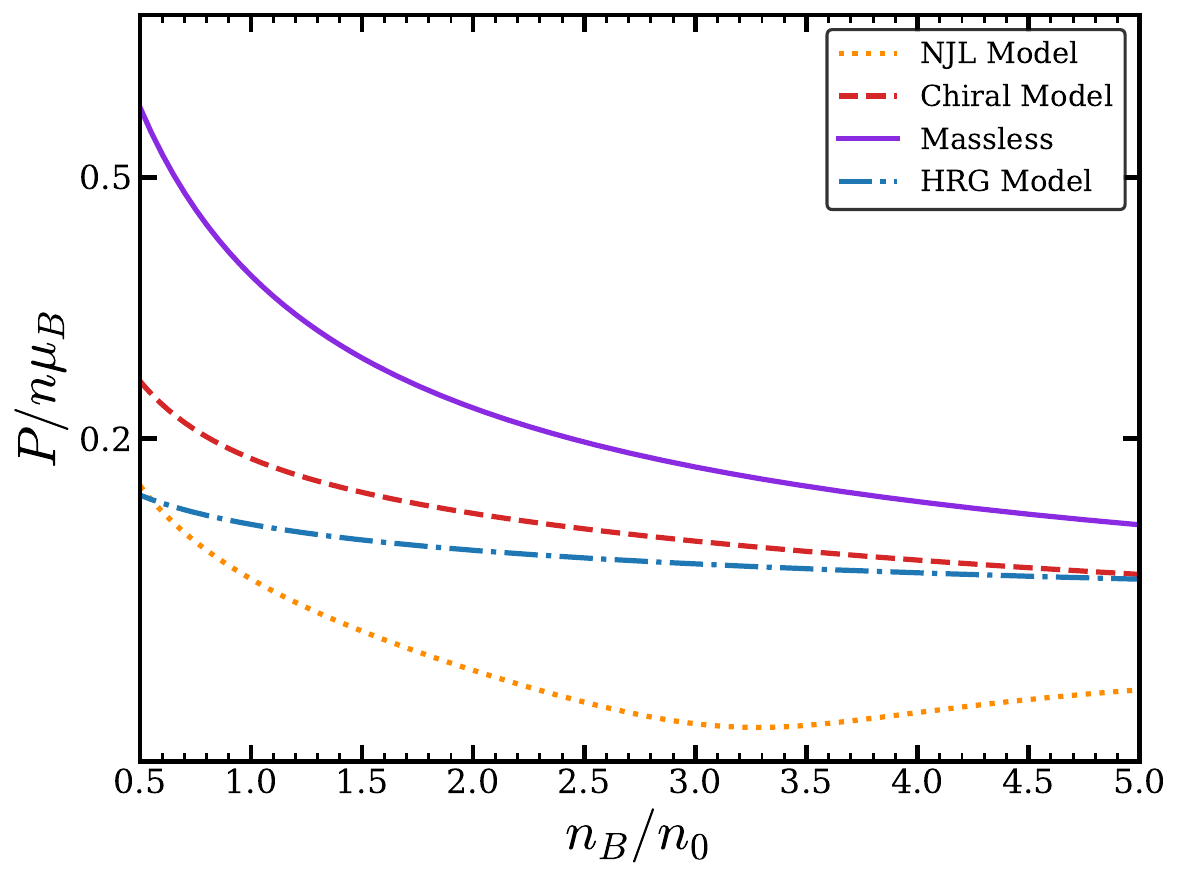}
	\end{subfigure}
	\begin{subfigure}{0.24\textwidth}
		\centering
		\includegraphics[width=\linewidth]{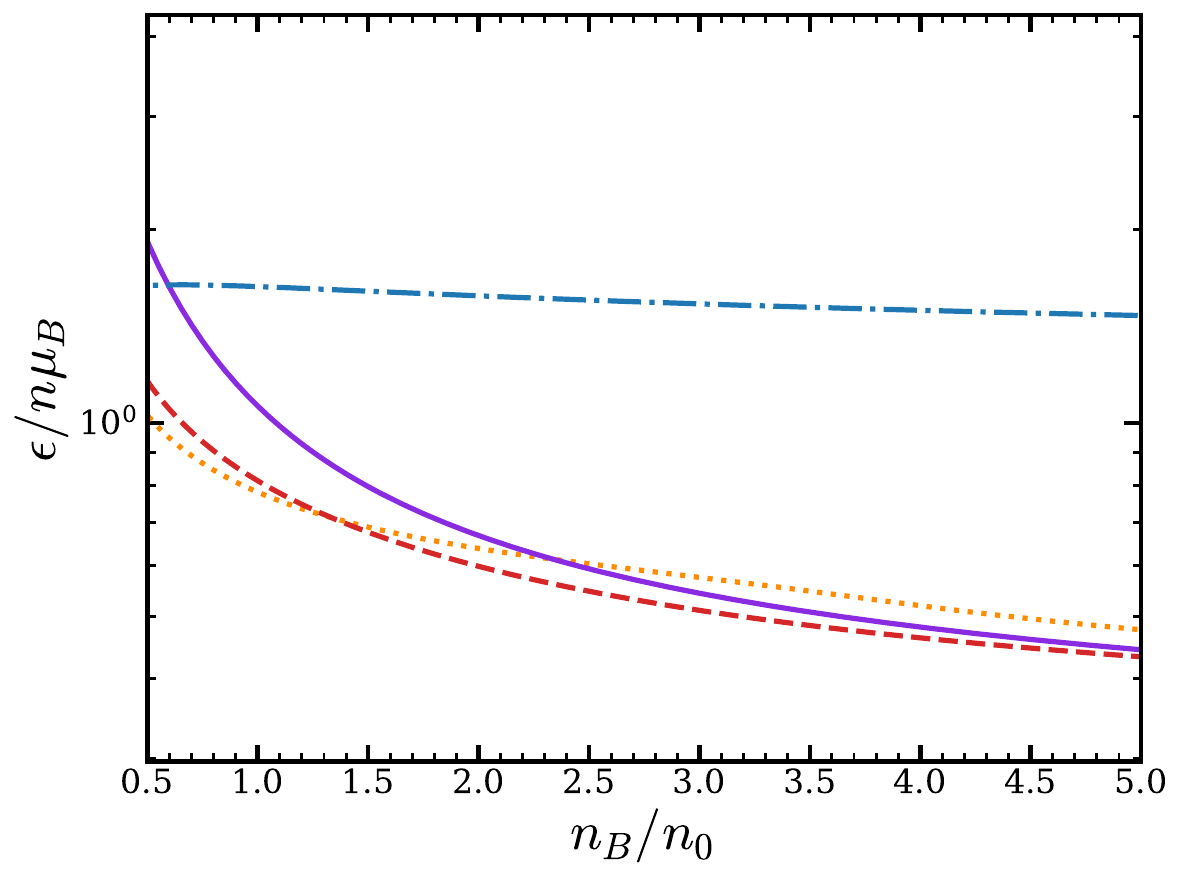}
	\end{subfigure}
	
	\vspace{0.5cm}
	
	\begin{subfigure}{0.24\textwidth}
		\centering
		\includegraphics[width=\linewidth]{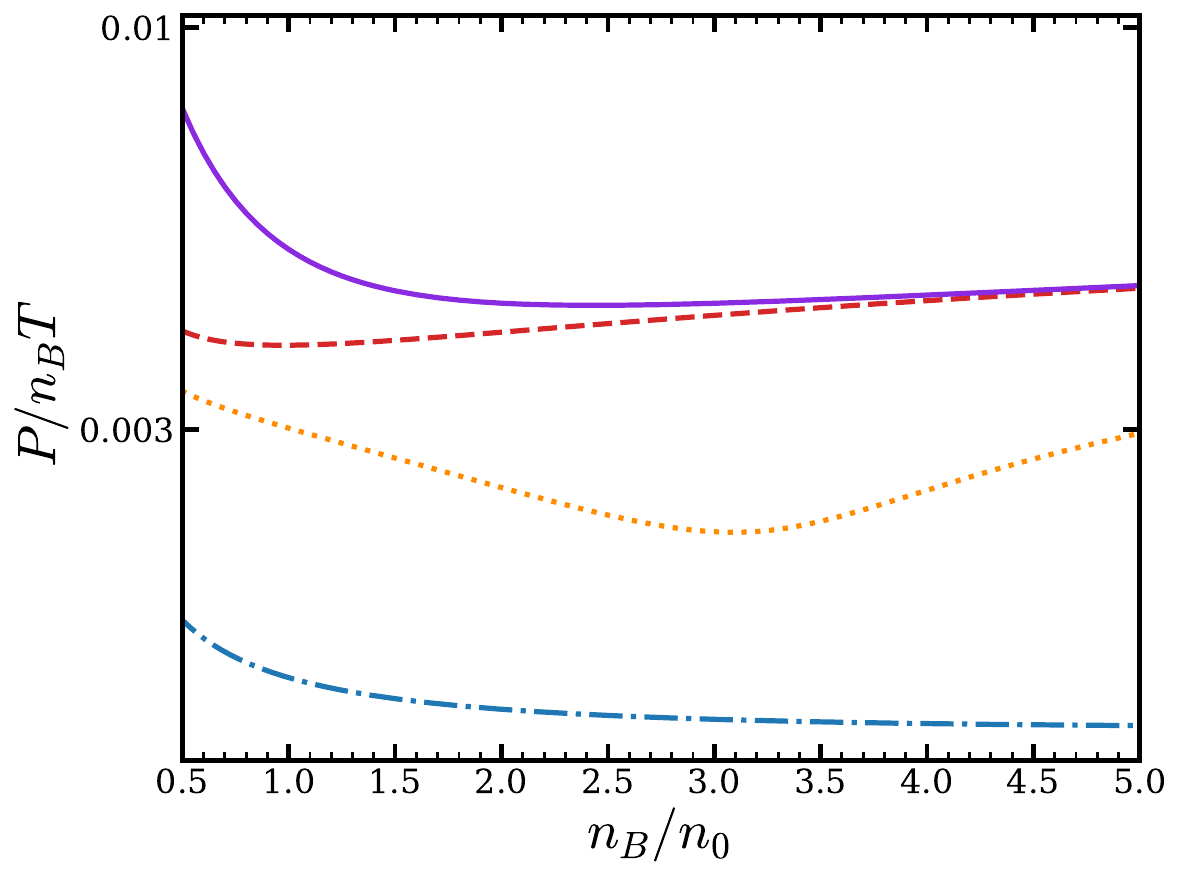}
	\end{subfigure}
	\begin{subfigure}{0.24\textwidth}
		\centering
		\includegraphics[width=\linewidth]{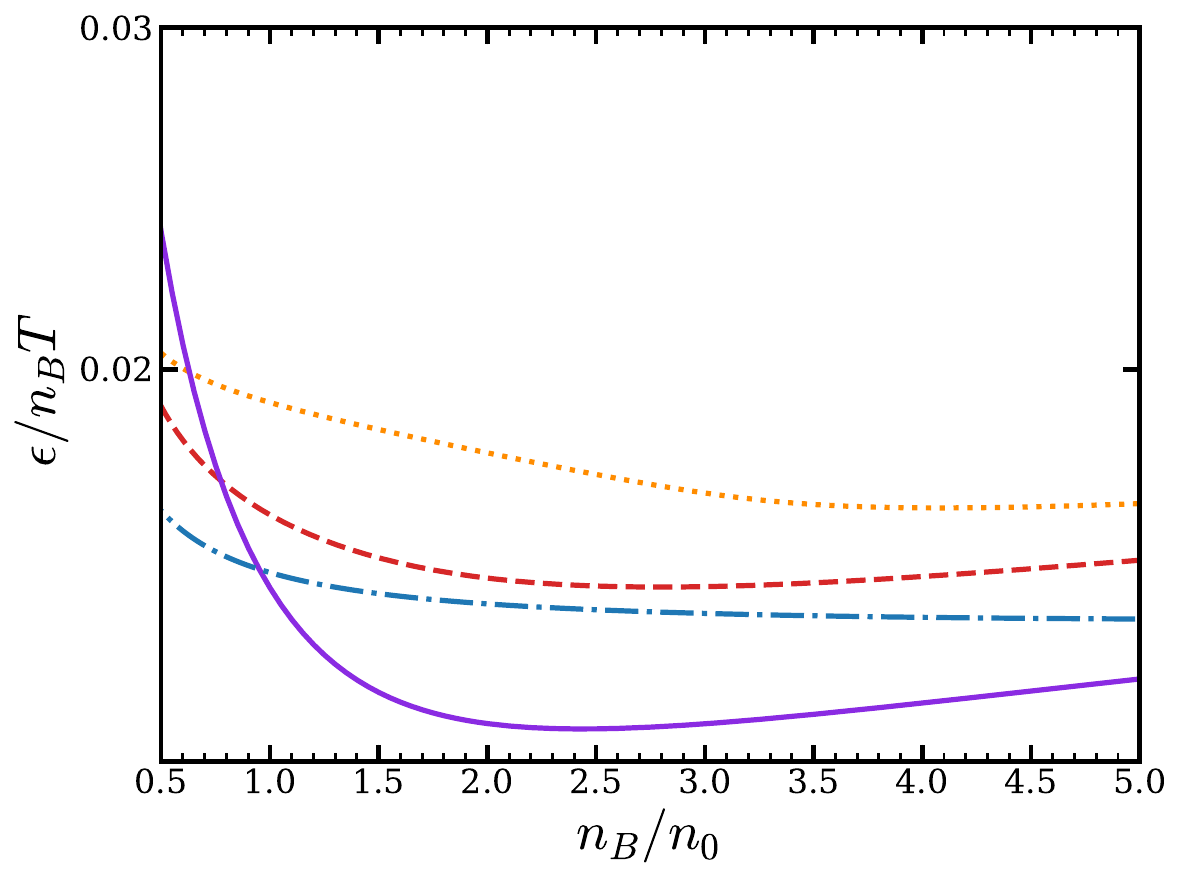}
	\end{subfigure}
	
	\caption{The dimensionless ratios $P/n\mu_B$, $\varepsilon/n\mu_B$, $P/n_BT$ and $\varepsilon/n_BT$  as a function of scaled baryon density $n_B/n_0$ in HRG, NJL and chiral effective models compared with the corresponding values for a massless three-flavor quark system.}
	\label{appendix2}
\end{figure}

\begin{figure}[htbp]
	\centering
	
	\begin{subfigure}{0.24\textwidth}
		\centering
		\includegraphics[width=\linewidth]{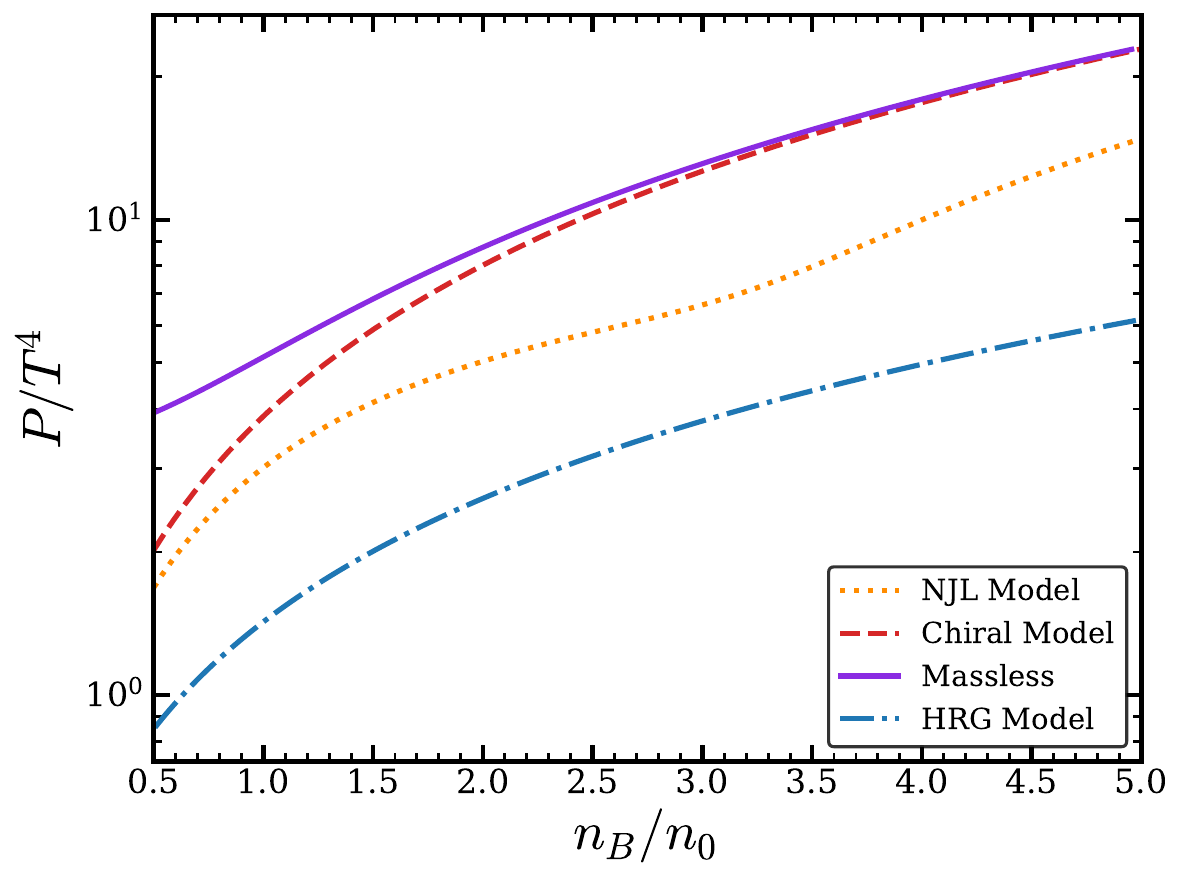}
	\end{subfigure}
	\begin{subfigure}{0.24\textwidth}
		\centering
		\includegraphics[width=\linewidth]{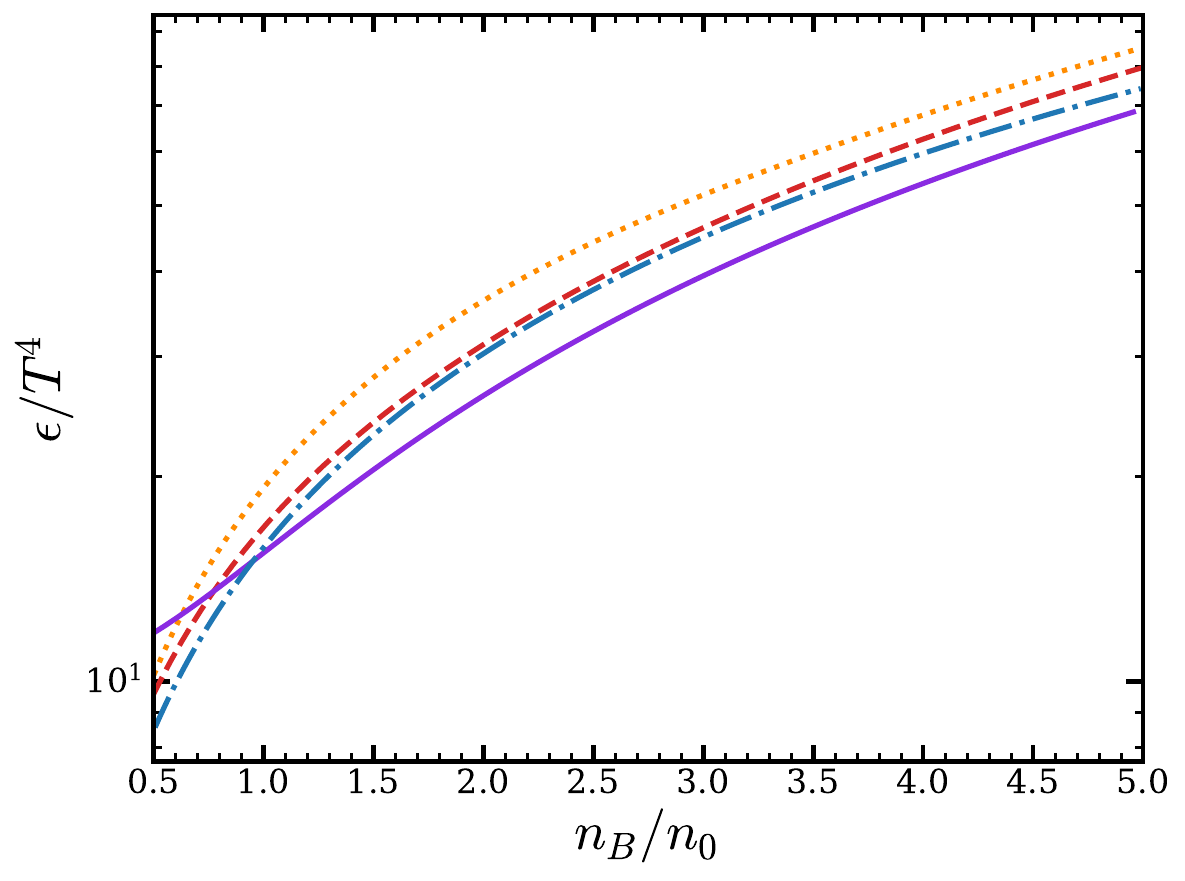}
	\end{subfigure}
	
	\vspace{0.5cm}
	
	\begin{subfigure}{0.24\textwidth}
		\centering
		\includegraphics[width=\linewidth]{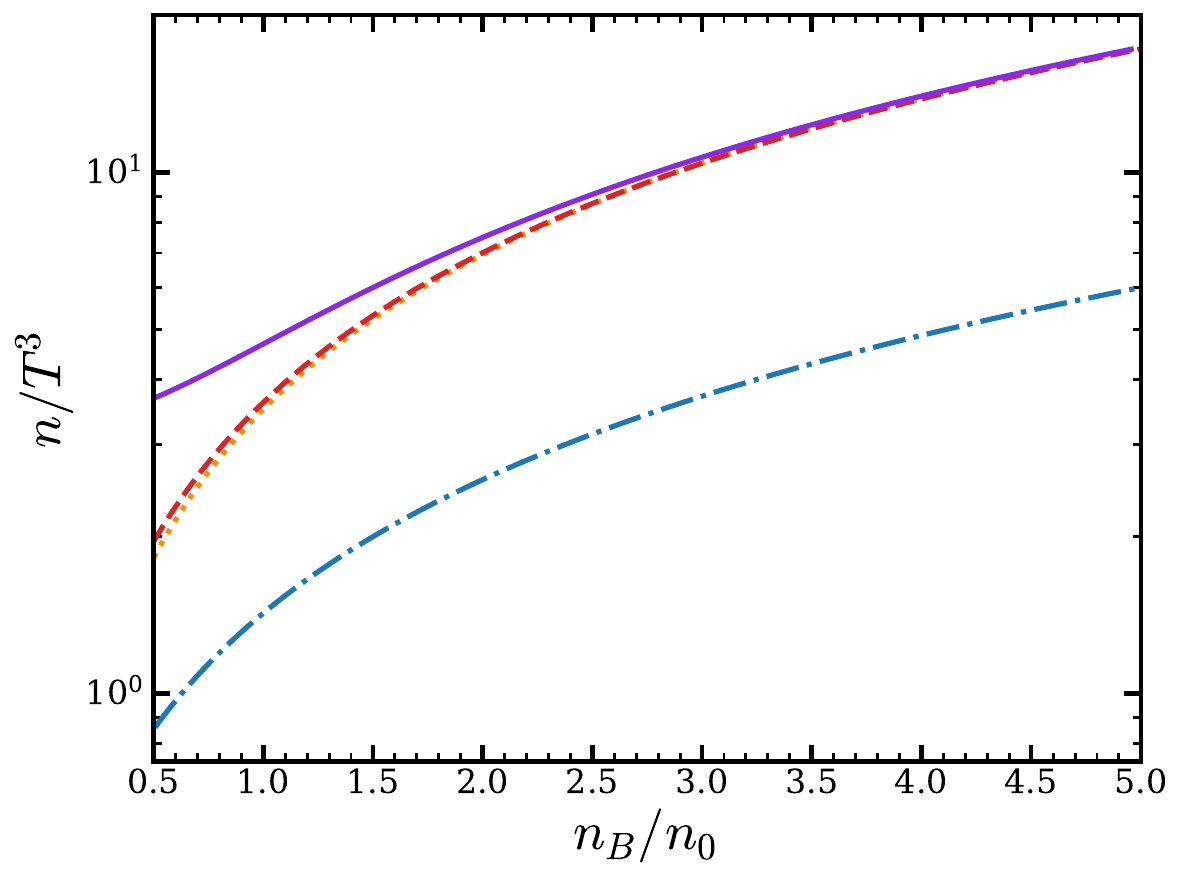}
	\end{subfigure}
	\begin{subfigure}{0.24\textwidth}
		\centering
		\includegraphics[width=\linewidth]{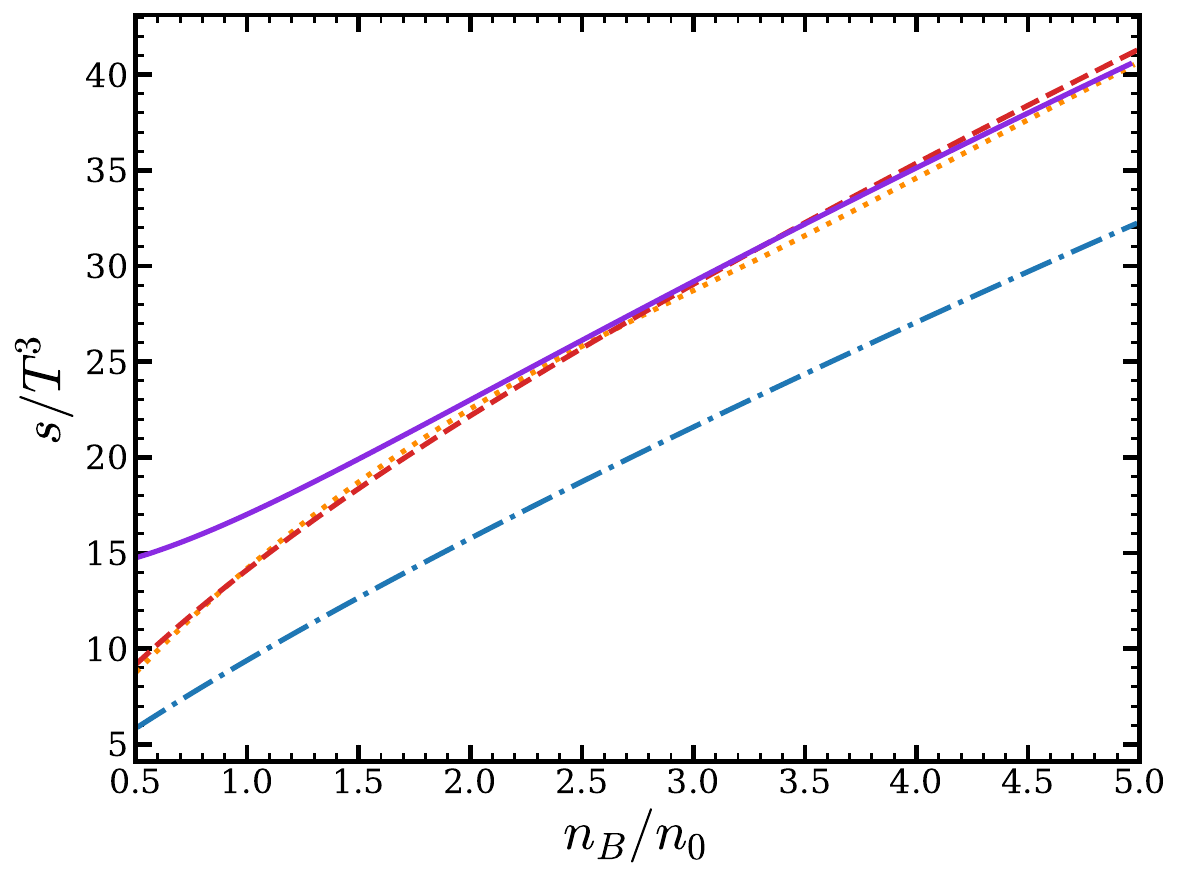}
	\end{subfigure}
	
	\caption{(Color Online) The dimensionless ratios $P/T^{4}$, $\varepsilon/T^{4}$, $n/T^{3}$ and $s/T^{3}$ as a function of scaled baryon density $n_B/n_0$ in HRG, NJL and chiral effective models compared with the corresponding values for a massless three-flavor quark system.}
	\label{appendix3}
\end{figure}

In this appendix, we provide supplementary figures that are complementary to the plots presented in Sec. \ref{sec:results}. In Fig.~\ref{appendix1}, the scaled total baryon density $n/n_0$ and net baryon density $n_B/n_0$ are presented as functions of $\mu_B/T$ in HRG, NJL, and chiral effective models as well as for a massless three-flavor quark system. This serves as an essential supplement to the Fig.~\ref{eta} for visualizing the variation of $\eta/s$ as a function of $n_{B}/n_{0}$.
Moreover, they also help to illustrate how other thermodynamic quantities and transport coefficients--presented in terms of $n_{B}/n_{0}$--vary as functions of $\mu_{B}/T$. We use the same color code as in the figures of Sec.~\ref{sec:results}, where blue dot-dashed, yellow dot, red dashed, and purple lines denote the results for HRG, NJL, chiral effective, and massless curves, respectively. The total baryon density $n$ increases slowly at lower $\mu_{B}/T=0.1-1$ and increases at a faster rate as $\mu_{B}/T$ further increases. In contrast, the net baryon density plot shows an exponential increase in $n_{B}$ starting from zero as $\mu_{B}$ increases from zero. As the $\mu_{B}/T$ increases the contribution of anti-baryons in both total baryon density $n=n_{\text baryons}+ n_{\text anti-baryons}$ and net baryon density $n=n_{\text baryons}- n_{\text anti-baryons}$ decreases as a result at very high $\mu_{B}/T$ we get $n\approx n_{B}$ for all the models. In Figs.~\ref{appendix2} and \ref{appendix3}, we provide the plots of some of the normalized thermodynamic variables with different normalization factors than those presented in Fig.~\ref{thermo_n}. As far as the equation of state is concerned, seeing the variation of $P(n_{B}, T)$ and $\varepsilon(n_{B}, T)$ at constant $T$ by normalizing them with $n\mu_{B}$ and $n_{B}T$ give complementary information in comparison to Fig.~\ref{thermo_n} bottom panel. Similarly, one may use an alternative normalization scheme at finite baryon density for the thermodynamic variables. Rather than normalizing with respect to different powers of $\mu_{B}$ (cf. Fig.~\ref{thermo_n}, top and middle panels), one can normalize with respect to different powers of $T$, as is done in Fig.~\ref{thermo_T}. We present such plots in Fig.~\ref{appendix3}. 

\bibliographystyle{unsrt}
\bibliography{ref}

\end{document}